\documentclass[secnumarabic,amssymb, nobibnotes, aps, prd]{revtex4-2}
\usepackage{graphicx}
\usepackage{amsmath}
\usepackage{subcaption}
\usepackage{xcolor}
\usepackage{multirow}

\begin{document}
	\title{Thermodynamics of Flat 4D Einstein-Gauss-Bonnet Black Hole with Rényi Entropy: An RPST-like formalism}
	\author{Amijit Bhattacharjee$^1$}
	
	\email{$rs_amijitbhattacharjee@dibru.ac.in$}
	\author{Prabwal Phukon$^{1,2}$}
	\email{prabwal@dibru.ac.in}	
	\affiliation{$1.$Department of Physics, Dibrugarh University, Dibrugarh, Assam,786004.\\$2.$Theoretical Physics Division, Centre for Atmospheric Studies, Dibrugarh University, Dibrugarh, Assam,786004.}
	\begin{abstract}
We investigate the thermodynamics of asymptotically flat black holes in four-dimensional Einstein-Gauss-Bonnet (4D-EGB) gravity using Rényi entropy as a non-extensive generalization of the Bekenstein-Hawking entropy. The resulting thermodynamic structure, formulated within a restricted phase space-like (RPST-like) framework, reveals a striking resemblance to the thermodynamics of AdS black holes in the standard RPST formalism. In particular, we identify a thermodynamic duality between the Rényi deformation parameter $\beta$ and a conjugate response potential $\zeta$, analogous to the central charge and chemical potential in holographic theories. An extensive thermodynamic analysis in both fixed charge-$(\tilde{Q})$ and fixed potential-$(\tilde{\Phi})$ ensembles reveal Van der Waals-like first-order phase transitions which is an unexpected feature for asymptotically flat black holes. Furthermore, through the formalism of geometrothermodynamics (GTD) and thermodynamic topology, It is shown that the Rényi modified flat black hole mimics, in both its thermodynamic topology and geometry, the features of  its counterparts in the 4D-EGB AdS black hole under RPST, reinforcing the structural similarity between these seemingly different systems. Our findings point to a deeper correspondence between non-extensive entropy and holographic thermodynamics, suggesting that Rényi entropy may serve as a natural bridge between flat-space black hole thermodynamics and AdS holography.
\end{abstract}

	\maketitle
	\section{Introduction}

The thermodynamics of black holes stands out as one of the most profound and conceptually rich intertwining between general relativity, quantum field theory, and statistical mechanics. Ever since the pioneering discoveries that black holes possess thermodynamic properties---most notably their entropy being proportional to the horizon area \cite{Bekenstein1973, Hawking1975} and that a temperature can be associated with their surface gravity at the event horizon \cite{Hawking1975}. Black holes have transcended their early interpretation as mere geometric solutions of Einstein’s equations. Instead, they are now regarded as genuine thermodynamic systems governed by laws remarkably analogous to those of conventional thermodynamics \cite{Bardeen1973}. Consequently, a wide array of theoretical frameworks has been developed to probe the thermodynamic behaviour of black holes under various physical settings, including those involving modified gravity theories \cite{Bamba2012, Nojiri2011}, quantum corrections \cite{Page2005, Carlip2014}, and holographic dualities \cite{Witten1998, Hubeny2010}. These efforts not only refine our understanding of black hole mechanics but also serve as critical probes into the quantum nature of space-time, potentially offering insights into a consistent theory of quantum gravity \cite{Rovelli2004, Ashtekar2005}.\\

The EPST (extended phase space thermodynamics) and the RPST (restricted phase space thermodynamics) are two of the most significant approaches that stand out among all such developments. In EPST \cite{Kastor2009, Kubiznak2012, Kubiznak2017}, the cosmological constant is promoted to be the thermodynamic pressure, and the black hole mass is interpreted as enthalpy, leading to rich thermodynamic phase structures including Van der Waals-like behaviour in black holes. RPST, on the other hand, stems from the holographic perspective rooted in the AdS/CFT correspondence \cite{Maldacena1998}. Unlike EPST where the cosmological constant is considered dynamical, RPST \cite{Cong2021, Visser2022} introduces a new pair of thermodynamic conjugates: the central charge $ C $, which quantifies the number of microscopic degrees of freedom in the dual conformal field theory, along with its conjugate chemical potential $ \mu $. In this formulation, the black hole mass is again regarded as internal energy, and the thermodynamic structure reflects holographic principles more directly. The thermodynamic variables $ P $ and $ V $ are constrained in RPST, which leads to a simplified first law of black hole thermodynamics:
\begin{equation}
\label{eq1}
dM = TdS + \Omega dJ + \tilde{\Phi} d\tilde{Q} + \mu dC,
\end{equation}

where $l$ is the AdS radius and by fixing it we eliminate any volume work thus simplifying the analysis of black hole thermodynamic processes.\\

While RPST has primarily been studied in asymptotically AdS space-times, its structural features suggest a more universal character that might extend beyond the AdS framework. Pertaining to which, numerous studies have been undertaken so as to extend the RPST framework to various non-AdS black holes  \cite{wang2022black,huang2024thermodynamics,kong2023restricted}  where the authors prescribe their RPST-like formalisms to investigate the thermodynamic structure of those non-AdS black holes. Our attempt though similar in spirit is unlike anything that has been done in the above mentioned works. We are here inspired by a completely different formalism as was given in \cite{hazarika2025rpst} where the authors employ Rényi entropy as a substitute for the Bekenstein-Hawking entropy of black holes. We here prescribe our RPST-like formalism in line with the work done in \cite{hazarika2025rpst} but we do so for black holes in modified theory of gravity i.e. the Einstein–Gauss–Bonnet (EGB) gravity \cite{Lovelock1971, Lanczos1938} which stands out with a compelling candidature due to its geometric elegance and its ability to encapsulate higher-curvature effects without compromising the structure of the field equations. EGB gravity is a natural extension of general relativity in higher dimensions that include quadratic curvature corrections while preserving second-order field equations.\\

 The study of EGB gravity offers a comprehensive framework for addressing various theoretical challenges in gravitational physics. In this context, a static, five-dimensional, spherically symmetric Hayward-like black hole solution was presented in \cite{kumar2020hayward}. Building upon this, the relationship between emission modes and the black hole’s temperature was explored in \cite{shahraeini2022radiation} via the tunneling of massless particles across the event horizon. However, in four-dimensional space-time, the Gauss--Bonnet term does not influence gravitational dynamics due to its topological nature. To address this, D.~Glavan and C.~Lin proposed a novel formulation of four-dimensional Einstein--Gauss--Bonnet (4D-EGB) gravity \cite{glavan2020einstein}, wherein the GB term is multiplied by a factor of $\frac{1}{(D - 4)}$ to bypass the restrictions of Lovelock’s theorem \cite{Lovelock1971} and avoid Ostrogradsky instability \cite{motohashi2015third}. In recent years, this modified theory has emerged as a valuable framework for investigating the physics and thermodynamics of black holes. Noteworthy studies include vacuum solutions for charged black holes and their thermodynamic characteristics \cite{fernandes2020charged}, analyses of black hole thermodynamics in the extended phase space \cite{wei2020extended,mansoori2021thermodynamic}, investigations involving charged black holes with entangled particle--antiparticle pairs \cite{bousder2021particle}, studies on stable circular orbits and black hole shadows \cite{guo2020innermost,vagnozzi2023horizon}, power spectra of Hawking radiation in de Sitter spacetimes \cite{zhang2020greybody}, assessments of black hole stability  \cite{zhang2020superradiance}, and non-linear charged planar black hole configurations \cite{bravo2022nonlinear,yang2020weak}.\\

In this work, we investigate the thermodynamics of flat 4D-EGB black holes \cite{glavan2020einstein} by adopting Rényi entropy as a substitute for the conventional Bekenstein–Hawking entropy for black holes. The Rényi entropy \cite{Renyi1970, Tsallis2013}, arising in generalized statistical mechanics, offers a framework for describing systems with non-extensive or long-range interactions, making it a compelling alternative for gravitational systems that may not conform to the assumptions of ordinary thermodynamics such as black holes. The expression for Rényi entropy is given by  \cite{Biro2013, Czinner2016}: 
\begin{equation}
S=\frac{1}{\lambda} ~ \ln[1+\lambda (S_{BH})]
\label{eq2}
\end{equation}
where $S_{BH}$ is the Bekenstein-Hawking entropy for black holes and $\lambda$ is the Rényi parameter. In the limit $\lambda \rightarrow 0$, $S \rightarrow S_{BH}$, thus recovering the standard thermodynamic description.\\

 Remarkably, we find that the resulting thermodynamic structure not only remains consistent, but closely resembles to that of the RPST framework originally formulated for AdS black holes. In particular, the emergence of conjugate pairs analogous to the central charge and chemical potential suggests that the holographic features of RPST may be encoded more generally in the thermodynamic description of gravity with modified entropy.
It is worth mentioning that recent investigations have revealed that flat black holes analyzed within the framework of Rényi entropy exhibit thermodynamic behaviour remarkably akin to that of asymptotically AdS black holes described under the Bekenstein-Hawking entropy paradigm. These similarities were first identified in \cite{Barzi1} and were subsequently explored across a range of contexts in \cite{Barzi2,proof3,proof4,proof5,proof6,proof7,proof8,proof9}. Notably, it has been shown in \cite{expanding} that the application of Rényi entropy in deriving the Friedmann equations allows the cosmological constant \(\Lambda\) to emerge as a function of the Rényi parameter \(\lambda\) i.e. $\Lambda \approx \pm \frac{3 \lambda \pi}{G}$, thereby eliminating the need for its ad-hoc inclusion in the Einstein–Hilbert action. This insight establishes a compelling and natural bridge between cosmology and non-extensive entropy. Moreover, we can easily convince ourselves with the help of very little algebra that such relationship can in fact quite naturally arise in terms of black holes. Where, to begin with, we may consider the metric function of a flat Schwarzschild black hole that can represent the simplest black hole solution possible, which is devoid of any other external parameters such as that of charge or rotation. The metric function of which can then be simply expressed as:
\begin{equation}
\label{eq3}
ds^2 = - \left(1 - \frac{2GM}{r} \right) dt^2 + \left(1 - \frac{2GM}{r} \right)^{-1} dr^2 + r^2 d\Omega
\end{equation}

Thus, from the metric function, the mass of the black hole as a function of event horizon radius $r_+$ is found to be:
\begin{equation}
\label{eq4}
M = \frac{r_+}{2G}
\end{equation}

Now, by using the expression for Rényi entropy, we calculate the expression for event horizon radius, which is given by:
\begin{equation}
\label{eq5}
r_+ = \frac{\sqrt{G} \sqrt{e^{\lambda S} - 1}}{\sqrt{\pi} \sqrt{\lambda}}
\end{equation}

Finally after substituting eqtn.(\ref{eq5}) into eqtn.(\ref{eq4}), we obtain:
\begin{equation}
\label{eq6}
M = \frac{\sqrt{e^{\lambda S} - 1}}{2 \sqrt{\pi} \sqrt{\lambda} \sqrt{G}}
\end{equation}

If we now consider $\lambda$ to be very small i.e. $(0< \lambda <1)$, we can expand eqtn.(\ref{eq6}) up to the first order in $\lambda$ and neglect the higher order terms. And therefore expanding eqtn.(\ref{eq6}) up to first order, we obtain:
\begin{equation}
\label{eq7}
M = \frac{\sqrt{S}}{2 \sqrt{\pi} \sqrt{G}} + \frac{\lambda (S)^{3/2}}{8 \sqrt{\pi} \sqrt{G}} + \mathcal{O}(\lambda^{3/2})
\end{equation}

 We then consider the metric for the Schwarzschild-AdS black hole, which is given as:
\begin{equation}
\label{eq8}
ds^2 = - \left(1 - \frac{2GM}{r} + \frac{r^2}{l^2} \right) dt^2 + \left(1 - \frac{2GM}{r} + \frac{r^2}{l^2} \right)^{-1} dr^2 + r^2 d\Omega
\end{equation}
where $l$ is the usual AdS length. From the metric function, the mass of the Schwarzschild-AdS black hole as a function of entropy $S$ can be written as:
\begin{equation}
\label{eq9}
M = \frac{\sqrt{S}}{2 \sqrt{\pi} \sqrt{G}} + \frac{\sqrt{G} S^{3/2}}{2 \pi^{3/2} l^2}
\end{equation}

By comparing eqtn.(\ref{eq9}) with the first two terms in eqtn.(\ref{eq7}), we obtain:
\begin{equation}
\label{eq10}
\lambda \approx \frac{4G}{\pi l^2}
\end{equation}

which as previously mentioned, is similar to the relation that was obtained in \cite{expanding}, where authors have shown a relation between $\lambda$ and cosmological constant $\Lambda$ as follows:
\begin{equation}
\label{eq11}
\Lambda \approx \pm \frac{3 \lambda \pi}{G}
\end{equation}
where, for $\Lambda= \pm \frac{3}{l^2}$ we get $\lambda \approx \frac{G}{\pi l^2}$. \\

Therefore, motivated by such developments, we propose the possibility of a deeper correspondence between the thermodynamic characteristics of black holes in asymptotically flat space-times governed by Rényi statistics and those in asymptotically AdS spacetimes governed by conventional Gibbs–Boltzmann statistics. Such structural parallels prompt the intriguing question of whether a direct and physical correlation exists between the cosmological constant \(\Lambda\) and the Rényi parameter \(\lambda\). In what follows, we aim to explore this potential connection in detail.\\

To provide further insight into the nature of this correspondence, we undertake a thermodynamic geometric analysis of the system using the formalism of geometrothermodynamics (GTD)\cite{Quevedo2007,Quevedo2008}. This approach, which constructs a Legendre-invariant Riemannian geometry on the thermodynamic phase space, allows one to detect critical behaviour and phase transitions through curvature singularities. The GTD metric is a Legendre invariant metric and therefore would not depend on any specific choice of thermodynamic potential. The phase transitions obtained from the specific heat capacity of the black hole are properly contained in the scalar curvature of the GTD metric, such that a curvature singularity in the GTD scalar `$R_{GTD}$'  would imply the occurrence of a phase transition. The general form of the GTD metric is given by \cite{{Soroushfar2016}} :
\begin{equation}
\label{eq12}
g = \left(E^{c} \frac{\partial{\varphi}}{\partial{E^{c}}} \right)
\left(\eta_{ab} \delta^{bc} \frac{\partial^2 \varphi}{\partial E^{c} \partial E^{d}} dE^{a} dE^{d}\right)
\end{equation}     
where `$\varphi$' is the thermodynamic potential and `$E^a$' is an extensive thermodynamic variable with $a=1,2,3....$..\\

Overall, this work demonstrates that the combination of Rényi entropy and 4D-EGB gravity naturally gives rise to a thermodynamic framework that parallels the RPST structure. Moreover, it shows that holographic-like behaviour and extended thermodynamic conjugacies may arise even in flat spacetimes, provided the entropy is appropriately generalized. The geometrothermodynamic perspective further enriches this picture by linking the phase structure to intrinsic geometric properties of the thermodynamic manifold.\\

In addition to that we also investigate the recently proposed topological method \cite{Wei2024,Wu2025,Wu2024,Wei2022PRL,Wu2023a,
Wu2023b,Wei2022PRD} in black hole thermodynamics, in the modified Rényi entropy framework so as to examine the potential modifications that it brings forth to the topology of the black hole. The most reliable method for studying the thermodynamic topology is to treat the black hole solutions as topological defects in their thermodynamic phase space. One can associate topological charges to these potential defects, based on which, the solutions can be classified into three distinct topological classes ($W=+1,0,-1$). By investigating the thermodynamic topology, we aim to observe the change in the topological class of the black hole solution due to the incorporation of the Renyi entropy as a substitute for the traditional Bekenstein-Hawking entropy.
The study of thermodynamic topology of black holes is primarily motivated by Duan’s $\phi$-mapping current theory \cite{Wei2024,Wu2025,Wu2024,Wei2022PRL}, which forms the key conceptual foundation. A concise overview of this framework is outlined below.\\

In order to explore the thermodynamic topology of the black hole, we use the off-shell free energy method, which is premised upon considering black hole solutions as topological defects within their thermodynamic context. This approach operates by analysing both the local and global topology by calculating winding numbers associated with these defects. These winding numbers can be used to categorize black holes based on their overall topological charge. More importantly, the thermal stability of a black hole can be correlated with the sign of its winding number. The core idea that lies at the heart of thermodynamic topology revolves around understanding these topological defects and their associated charges.  We discuss the essential mathematical procedures involved in investigating thermodynamic topology as given below.\\

A vector field in black hole thermodynamics is derived from the generalized off-shell free energy and the expression for the off-shell free energy of a black hole with arbitrary mass is given by \cite{Wei2024,Wu2025}
\begin{equation}
\mathcal{F} = E - \frac{S}{\tau},
\label{eqF} 
\end{equation}
where $E$ is the energy (equivalent to the mass $M$) and $S$ denotes the entropy of the black hole. The parameter $\tau$, which represents the time scale is allowed to vary freely. Now, in order to make use of this generalized free energy a vector field $\phi$ is defined as \cite{Wei2024,Wu2025}
\begin{equation}
 \vec{\phi} = (\phi^{r}, \phi^{\Theta}) = \left( \frac{\partial \mathcal{F}}{\partial r_{+}}, \, -\cot \Theta \, \csc \Theta \right).
\label{eqT}
\end{equation}
The zero points of this vector field are given as: $(\tau, \Theta) = \left( \frac{1}{T}, \frac{\pi}{2} \right)$ where $T$ represents the equilibrium temperature of the cavity that surrounds the black hole (i.e. on-shell temperature). The points at which a vector field either diverges or becomes ill defined hold significant physical meaning. In our framework, such points correspond to the zeros or defects of the vector field, which characterize black hole solutions. Thus, black holes may be interpreted as topological defects of the constructed vector field, with each solution carrying an associated topological charge. We use Duan’s $\phi$-mapping technique to determine the associated topological charge. We calculate the unit vector $n$ of the field in Eq. (\ref{eqT}), which are given as:
\begin{equation}
n^{1} = \frac{\phi^{r}}{\sqrt{(\phi^{r})^{2} + (\phi^{\Theta})^{2}}}, 
\qquad
n^{2} = \frac{\phi^{\Theta}}{\sqrt{(\phi^{r})^{2} + (\phi^{\Theta})^{2}}}.
\label{eqn}
\end{equation}
The vector $n^{a}$ must satisfy two key conditions $n^{a}n^{a} = 1$, and $n^{a}\partial_{\nu}n^{a} = 0$, where $\partial_{\nu} = \frac{\partial}{\partial x^{\nu}}$ and $\mu,\nu,\rho = 0,1,2$. We construct a topological current $j^{\mu}$ in the coordinate space $x^{\nu} = \{t, r_{+}, \theta\}$ defined by
\begin{equation}
j^\mu=\frac{1}{2\pi}\epsilon^{\mu \nu \rho}\epsilon_{ab}\partial_\nu n^a \partial_\rho n^b.
\end{equation}
It can be directly verified that the current $j^\mu$ is conserved, i.e.  
\begin{equation}
	\partial_\mu j^\mu=0.
\end{equation}
Employing the Jacobian tensor $\epsilon^{ab} J^\mu \!\left(\frac{\phi}{x}\right)=\epsilon^{\mu\nu\rho} \partial_\nu \phi^a \partial_\rho \phi^b$ together with the two-dimensional Laplacian Green’s function $\Delta_{\phi^a}\ln ||\phi||=2\pi \delta^2(\phi)$, one can rewrite the current in the compact form \cite{Wei2024,Wu2025}  
\begin{equation}
	j^\mu=\delta^2(\phi) J^\mu\!\left(\frac{\phi}{x}\right).
\end{equation}
This expression indicates that $j^\mu$ takes non-zero values only at the zeros of the vector field $\phi^a(x^i)$, where each solution is denoted by $\vec{x}=\vec{z}_i$. Applying the $\delta$-function theory \cite{Schouten1951}, the temporal component of the current becomes  
\begin{equation}
	j^0=\sum_{i=1}^N \beta_i \eta_i \delta^2 (\vec{x}-\vec{z}_i).
\end{equation}
Here, $\beta_i$ is the Hopf index, which counts the number of times $\phi^a$ winds around the origin in the $\phi$-space when $x^\mu$ encircles the zero point $z_i$. The Brouwer degree $\eta_i$ is defined as $\eta_i=\text{sign}\!\left( J^0(\phi/x)_{z_i} \right)=\pm 1$. Consequently, within a chosen parameter domain $\Sigma$, the total topological charge is given by  
\begin{equation}
	W=\int_\Sigma j^0 d^2x=\sum_{i=1}^N \beta_i \eta_i=\sum_{i=1}^N w_i,
\end{equation}
where $w_i$ denotes the winding number associated with the $i$-th zero of $\phi$.\\

The remainder of the paper is organized as follows. In Section~\ref{sec:RenyiEGB}, we present the flat 4D-EGB black hole solution and transform it with the Rényi entropy framework. Section~\ref{sec:thermo} develops the black hole thermodynamics under this modified entropy and elucidates its resemblance to the RPST formalism both in fixed charge-$(\tilde{Q})$ and fixed potential-$(\tilde{\Phi})$ ensembles. Section~\ref{sec:GTD} is devoted to the geometrothermodynamic  along with the thermodynamic topological analysis of the black hole system. We conclude with a summary and discussion in Section~\ref{sec:conclusion}.

\section{Flat 4D-EGB Black Hole in the Rényi Entropy Formalism}
\label{sec:RenyiEGB}
The action for the Einstein-Gauss-Bonnet gravity in a d-dimensional space-time with a negative cosmological constant is given by \cite{fernandes2020charged}: 
\begin{equation}
	\label{eq13}
I=\frac{1}{16 \pi G} \int d^d x\left(R - 2\Lambda-F_{\mu \nu} F^{\mu \nu} +\frac{\alpha}{d-4} \mathcal{G}\right),
\end{equation}
where the Gauss-Bonnet(GB) term writes  
\begin{equation}
\mathcal{G}=R_{\mu \nu \rho \sigma} R^{\mu \nu \rho \sigma}-4 R_{\mu \nu} R^{\mu \nu}+R^2,
\end{equation}
and $\alpha$ here is the Gauss-Bonnet coupling parameter with dimension of length squared which represents ultraviolet
corrections to the Einstein theory, $G$ is the well known Newton’s gravitational constant and the cosmological constant is given by  
\begin{equation}
\Lambda=-\frac{(d-1)(d-2)}{2 l^2}.
\end{equation}
 The AdS radius is given by $l$ and $  F_{\mu \nu}=\partial_\mu A_\nu-\partial_\nu A_\mu  $ is the electromagnetic field strength tensor. The black hole metric, by adopting the limit $d\longrightarrow 4$, is given by \cite{glavan2020einstein,FernandesReview2022}: 

\begin{equation}
d s^2=-f(r) d t^2+\frac{1}{f(r)} d r^2+r^2\left(d \theta^2+\sin ^2 \theta d \phi^2\right),
\end{equation}
where the metric function is:
\begin{equation}
\label{eq14}
f(r)=1+\frac{r^2}{2 \alpha}\left(1-\sqrt{1+4 \alpha\left(\frac{2 M G}{r^3}-\frac{Q^2 G}{r^4}-\frac{1}{l^2}\right)} \right),
\end{equation}
where M and Q are the mass and electric charge for the black hole respectively. Solving the lapse function i.e. $f(r)=0$, we find two horizons:  the event horizon $r_+$ and the Cauchy horizon $r_-$. Using eqtn.(\ref{eq14}) we write,  $f(r_{+}) = 0$, and the black hole's mass is therefore given by:

\begin{equation}
\label{eq15}
M=\frac{r_+}{2G} + \frac{\alpha}{2\, r_+ G} + \frac{r_+^3}{2\, l^2 G} + \frac{ Q^2}{2\, r_+ }
\end{equation}
In order to get a flat 4D-EGB black hole we take $\Lambda=-\frac{(d-1)(d-2)}{2 l^2}=0$ and therefore the mass for the flat 4D-EGB black hole is obtained as:
\begin{equation}
\label{eq16}
M=\frac{r_+}{2G} + \frac{\alpha}{2\, r_+ G} +  \frac{ Q^2}{2\, r_+ }
\end{equation}
As was previously mentioned we are here interested in combining the flat 4D-EGB gravity with the Renyi entropy framework in order to establish how its thermodynamic behaviour can mimic a RPST like structure. So in order to do so, we write the Renyi entropy, $S$ in terms of the Bekenstein-Hawking entropy, $S_{BH}$ to finally compute the horizon radius, $r_+$ and replace it in the mass for the flat 4D-EGB black hole. The Renyi entropy can be written as:
\begin{equation}
S=\frac{1}{\lambda} ~ ln[1+\lambda (S_{BH})],~~~~~~~~S_{BH}=\frac{\pi r_+^2}{G}
\label{eq17}
\end{equation}
and therefore $r_+$ is given by:
$$r_+ = \sqrt{ \frac{G}{\pi \lambda} \left( e^{\lambda S} - 1 \right) }$$
where, for computational purposes we replace the renyi parameter, $\lambda$ with $\frac{1}{\beta}$ to finally obtain:
\begin{equation}
r_+ = \sqrt{ \frac{G \beta}{\pi} \left( e^{S/\beta} - 1 \right) }
\label{eq18}
\end{equation}
where, the newly introduced parameter $\beta$ is thereby just the inverse of the Renyi parameter $\lambda$. We thus replace $r_+$ from eqtn.(\ref{eq18}) to eqtn.(\ref{eq16}) in order to obtain the Renyi modified mass of the 4D-EGB flat black hole, which is given by:
\begin{equation}
M = \frac{ \beta \left( e^{S/\beta} - 1 \right) + \pi \alpha + \pi Q^2 }{ 2 \sqrt{ \pi G \beta \left( e^{S/\beta} - 1 \right) } }
\label{eq19}
\end{equation}
In frameworks governed by Rényi entropy, the parameter $\beta$ can be interpreted as quantifying the effective number of degrees of freedom or the richness of accessible information. This role is analogous to that of the central charge $C$ in the AdS-RPST formalism, where $C$ characterizes either the extent of the dual bulk space-time or the density of states in the boundary conformal field theory (CFT). Motivated by this parallel, we propose a formulation of the first law of black hole thermodynamics inspired by the RPST framework, extending it to non-AdS (asymptotically flat) black holes which is given as:
 \begin{equation}
dM = TdS +  \Phi dQ + \mathcal{A} d\alpha + \zeta d\beta,
\end{equation}  
where $\mathcal{A}=\frac{\partial M}{\partial \alpha}$ is the conjugate potential for the GB parameter $\alpha$ and, $\zeta = \frac{\partial M}{\partial \beta}$ is the conjugate potential to $\beta$. The parameter $\zeta$ can be considered to be the "response potential" that quantify the sensitivity of the thermodynamic system to changes in $\beta$, similar to the chemical potential $\mu$ in traditional thermodynamics. Physically, $\zeta$ is a description of how variations in the scaling properties of the thermodynamic system which, governed by $\beta$ can influence macroscopic quantities such as the mass $M$ or entropy $S$ of the black hole. Together, $\beta$ and $\zeta$ form a thermodynamic duality, with $\beta$ representing the structure of entropy and degrees of freedom and $\zeta$ being its conjugate potential which is responsible for the system's response to the changes in $\beta$. This duality parallels with that of the $C$-$\mu$ duality found in traditional AdS RPST, where, the parameter $C$ represents the density of states or the system's informational content, while $\mu$ is responsible for thermodynamic modifications which arise from variations in the composition of the state. In this extended framework, the parameter $\beta$ characterizes the possible deviations from conventional statistical mechanics thus effectively capturing the system's sensitivity to scaling and its degree of non-trivial behaviour. Meanwhile, $\zeta$  regulates how such deviations influence macroscopic thermodynamic properties and therefore offering insights into the nature of underlying interactions being either predominantly repulsive or attractive. Although the entropy expression adheres to the non-extensive Rényi formulation, the thermodynamic structure developed in the spirit of the RPST framework preserves extensivity. \\

As was previously stressed upon in eqtn.(\ref{eq11}) we here rewrite the relation obtained between the cosmological constant $\Lambda$ and the Renyi parameter $\lambda$ as:
$$\Lambda \approx \pm \frac{3 \lambda \pi}{G}$$
which after replacing $\Lambda=-\frac{3}{ l^2}$ and $\lambda= \frac{1}{\beta}$ turns into:
\begin{equation}
\beta \approx \pm \frac{\pi l^2}{G}
\label{eq20}
\end{equation}
Now, inspired by the relation in eqtn.(\ref{eq20}) we use a rescaling constant $\kappa$ to rewrite the Newton's constant $G$, rescaled GB parameter $\tilde{\alpha}$ and rescaled electric charge $\tilde{Q}$ as:
	
	\begin{equation}
	\tilde{Q} \to \frac{\kappa ~Q}{\sqrt{G}}, ~~~~~~G\to \frac{\kappa ^2}{ \beta}~~~~~~\text{and}~~~~~~\tilde{\alpha} \to \frac{\alpha}{{\kappa}^2} 
	\label{eq21}
	\end{equation}

 	Using eq.(\ref{eq21}) in eq.(\ref{eq19}), the mass is rewritten as: 
	\begin{equation}
	M=\frac{{\beta}^2 \left( e^{S/\beta} - 1 \right) + \pi {\beta}^2 \tilde{\alpha} + \pi  \tilde{Q}^2}{2 \sqrt{\pi } \kappa  \beta  \sqrt{e^{S/\beta }-1}}
	\end{equation}
	And finally the rescaled mass $\tilde{M}=\kappa M$, is given by:
	\begin{equation}
	\tilde{M}=\frac{{\beta}^2 \left( e^{S/\beta} - 1 \right) + \pi {\beta}^2 \tilde{\alpha} + \pi  \tilde{Q}^2}{2 \sqrt{\pi } \beta  \sqrt{e^{S/\beta }-1}}
	\label{eq22}
	\end{equation}
	
If $S$, $\tilde{Q}$, $\mathcal{\tilde{A}}$ and $\beta$ are rescaled as $S \to \gamma S$, $\tilde{Q} \to \gamma \tilde{Q}$, $\mathcal{\tilde{A}} \to \gamma \mathcal{\tilde{A}}$ and $\beta \to \gamma \beta$, then eqtn.(\ref{eq22}) implies  $\tilde{M} \to \gamma \tilde{M}$  which proves the first order homogeneity of
$\tilde{M}$ as was previously obtained in \cite{Ladghami2023} for Einstein-Gauss-Bonnet-AdS black hole in the usual Bekenstein-Hawking entropy paradigm. 
\section{Thermodynamic Analysis of Rényi modified Flat 4D-EGB Black Hole in the RPST like formalism}
\label{sec:thermo}
Using the expression for mass in eqtn.(\ref{eq22}), the other quantities are computed as given below:
\begin{equation}
T=\frac{e^{S/\beta} \left( \left( -1 + e^{S/\beta} \right) \beta^2 - \pi \left( \tilde{Q}^2 + \tilde{\alpha} \beta^2 \right) \right)}{4 \left( -1 + e^{S/\beta} \right)^{3/2} \sqrt{\pi} \, \beta^2}
\label{eq23}
\end{equation}
\begin{equation}
\tilde{\phi}=\frac{\sqrt{\pi } \tilde{Q}}{\beta  \sqrt{e^{S/\beta }-1}}
\label{eq24}
\end{equation}
\begin{equation}
\zeta=-\frac{e^{\frac{2S}{\beta}} (S - 2\beta) \beta^2 
    + e^{\frac{S}{\beta}} \left( \pi \tilde{Q}^2 (S - 2\beta) + (S - 4\beta) \beta^2 + \pi \tilde{\alpha} \beta^2 (S + 2\beta) \right) 
    + 2 \left( \beta^3 + \pi \beta (\tilde{Q}^2 - \tilde{\alpha} \beta^2) \right)}{4 \left( -1 + e^{\frac{S}{\beta}} \right)^{3/2} \sqrt{\pi} \,\beta^3}
\label{eq25}
\end{equation}
The eqtns.(\ref{eq23})-(\ref{eq25}) distinctly show the zeroth order homogeneity of all these above mentioned quantities. Therefore now, by dint of the mass of the black hole in eqtn.(\ref{eq22}), the first law of black hole thermodynamics for the Renyi modified flat 4D-EGB black hole can be written as: 
\begin{equation}
d\tilde{M}=T~d S+ \tilde{\phi}~d \tilde{Q}+\mathcal{\tilde{A}}~d\tilde{\alpha} +\zeta ~d \lambda
\end{equation}
where $\tilde{Q}$, $\tilde{\phi}$, $\tilde{\alpha}$ and $\mathcal{\tilde{A}}$  are the rescaled electrical charge, electric potential, Einstein-Gauss bonnet coupling parameter and its conjugate potential respectively. We have here introduced two new quantities namely: $\beta$ and $\zeta$ which are the Renyi deformation parameter and response potential respectively. It is quite trivial to show that the following Euler relation holds: 
\begin{equation}
E=T~S+\tilde{\phi}~\tilde{Q}+\mathcal{\tilde{A}}~\tilde{\alpha}+ \zeta~\beta
\end{equation}
\subsection{Thermodynamic process in the fixed ($\tilde{Q}$) ensemble}

In this section, we will look into the various allowed thermodynamic processes of the black hole namely: the $T-S$ and the $F-T$ processes. In order to obtain the critical points, we solve the following pair of equations which are given by :
\begin{equation}
\frac{d T}{d S}=0,~~~~~~~~~~~~\frac{d^2 T}{d S^2}=0
\end{equation}
Solving the above mentioned equations we obtain the critical value of $S$ and $\tilde{Q}$ which are given as:
\begin{equation}
S_C=\beta~  ln \left(2 \left(\sqrt{3}-1\right)\right) ~~~~~~~~~~\tilde{Q}_C= \beta~ \sqrt{\frac{7-4 \sqrt{3}}{\pi } - \tilde{\alpha} } 
\label{eq26}
\end{equation}
By substituting the critical values from eqtn.\ref{eq26} in the expression for $T$ in eqtn.\ref{eq23}, the critical temperature, $ T_C$ is found to be $ T_C=0.256236$, using which the relative temperature, $t$ can be written as:
\begin{equation}
t = \frac{
2^{-2+s}
\left(-1 + \sqrt{3}\right)^{-1+s}
\left(-3 + 2\sqrt{3}\right)^{3/2}
\left(-1 + 2\left(-1 + \sqrt{3}\right)\right)^s
- \pi \tilde{\alpha}
+ q^2 \left(-7 + 4\sqrt{3} + \pi \tilde{\alpha}\right)
}{
\left(-5 + 3\sqrt{3}\right)
\left(-1 + \left(2\left(-1 + \sqrt{3}\right)\right)^s\right)^{ 3/2}}
\end{equation}

where 
\begin{equation}
t=\frac{T}{T_C},~~~~s=\frac{S}{S_C},~~~~~q=\frac{\tilde{Q}}{\tilde{Q}_C}
\end{equation}
The rescaled Helmholtz free energy $f=F/F_C$ is obtained to be
\begin{equation}
f = \frac{
3 \left(2\left(-1 + 2\left(-1 + \sqrt{3}\right)\right)\right)^s 
\left(-1 + 2\left(-1 + \sqrt{3}\right)\right)^s 
+ \pi \tilde{\alpha} 
- q^2 \left(-7 + 4\sqrt{3} + \pi \tilde{\alpha}\right)-d
}{ 
\left(4\sqrt{-3} + 2\sqrt{3}\right)
\left(-1 + \left(2\left(-1 + \sqrt{3}\right)\right)^s\right)^{ 3/2}
\left(\sqrt{3} - 2 \log\left[2\left(-1 + \sqrt{3}\right)\right]\right)
}
\end{equation}

where, $d=\left(2\left(-1 + \sqrt{3}\right)\right)^s 
\left(-1 + 2\left(-1 + \sqrt{3}\right)\right)^s 
\left(
- \pi \tilde{\alpha} 
+ q^2 \left(-7 + 4\sqrt{3} + \pi \tilde{\alpha}\right)
\log\left[2\left(-1 + \sqrt{3}\right)\right]
\right)$.\\

\begin{figure}[ht]
\begin{center}
\includegraphics[width=.52\textwidth]{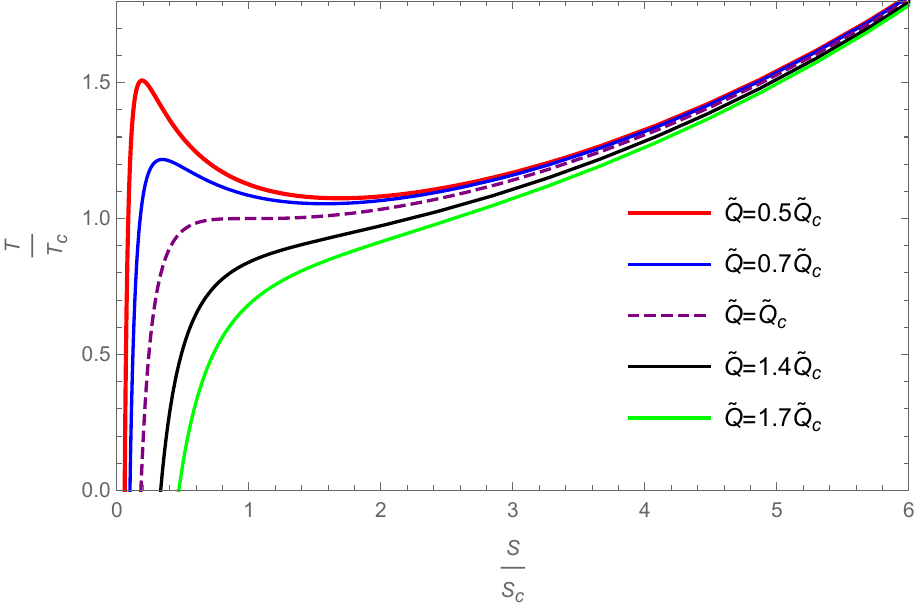}\vspace{3pt}
\includegraphics[width=.48\textwidth]{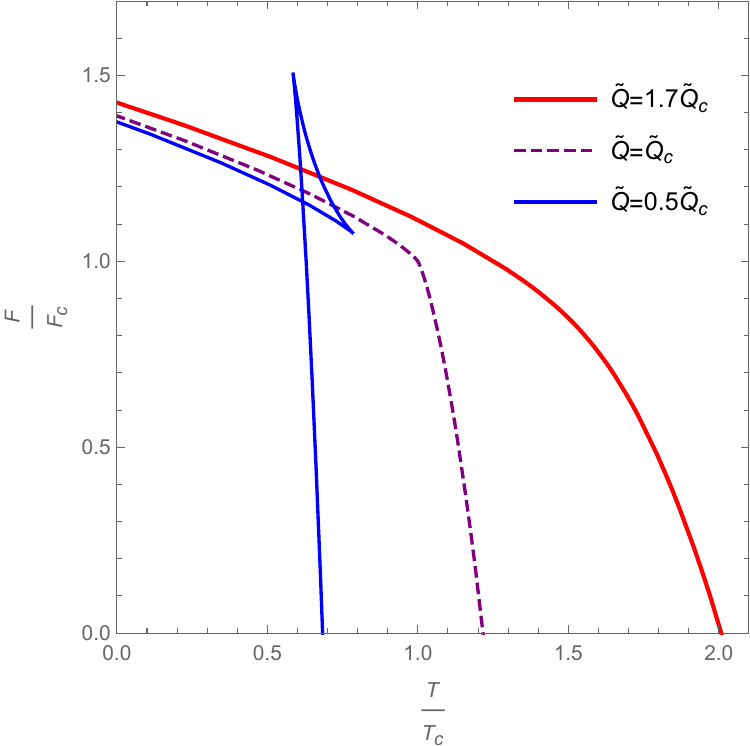}
\caption{$T-S$ and $F-T$ curves in the iso-$e$-charge processes for $\tilde{\alpha}=0.002$ and $\beta=\frac{1}{\lambda}=83.33$.}\label{fig1}
\end{center}
\end{figure}
 FIG. \ref{fig1} here illustrates both the $T-S$ and $F-T$ curves for the iso-e-charge process for the Renyi modidfied flat 4D-EGB black hole. The plots quite distinctly show that below the critical value of $\tilde{Q}_C$ both the $T-S$ and $F-T$ curves show non-trivial behaviour where, the $T-S$ curve exhibits non-monotonic behaviour below $\tilde{Q}_C$ whereas the $F-T$ curve displays a swallowtail behaviour below the critical value of $\tilde{Q}_C$. Such behaviour is an indication for the possible Van der Waals-like first-order phase equilibrium in the iso-e-charge processes when $0 < \tilde{Q} < \tilde{Q}_C$. However at the critical point $\tilde{Q} = \tilde{Q}_C$ we see that the second order phase transition is observed as represented by the dashed solid curve in both the figures.\\

It is noteworthy to emphasize upon the fact that the first-order Van der Waals-like phase transition cannot be found in the thermodynamics of non-AdS black holes. In flat charged black holes, only Davies-type phase transitions are a possibility. However, we find that with the employment of our RPST-like formalism, a first-order Van der Waals-like phase transition emerges, similar to what can be observed in the RPST framework for various charged AdS black holes.\\

\begin{figure}[ht]
\begin{center}
\includegraphics[width=.48\textwidth]{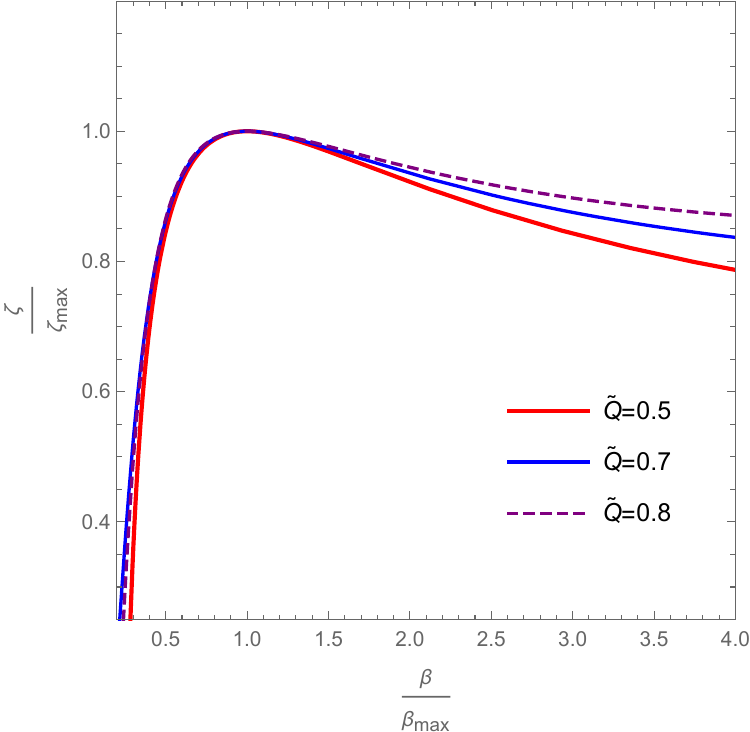}
\caption{$\zeta-\beta$ curves in the iso-$e$-charge processes for $\tilde{\alpha}=0.002$ and $S=1$.}\label{fig2}
\end{center}
\end{figure}

 We then analyze the $\zeta-\beta$ process in detail for the iso-e-charge process. We first plot the expression for $\zeta$ as a function of $\beta$ using eqtn.(\ref{eq25}), keeping $\tilde{Q}$ and $S$ fixed. The corresponding plot is shown as  Fig.\ref{fig2}. The plot reveals that its behaviour is remarkably akin to the $\mu-C$ process in the AdS RPST formalism for the non-modified EGB black hole.

\subsection{Thermodynamic process in the fixed ($\tilde{\Phi}$) ensemble}

In this section, we will look into the various allowed thermodynamic processes of the black hole namely: the $T-S$ and the $F-T$ processes in the fixed $\tilde{\Phi}$ ensemble. In order to obtain the critical points, we first replace $\tilde{Q}$ in terms of $\tilde{\Phi}$ by the dint of eqtn.\ref{eq24} and then we proceed to solve the following pair of equations which are given by :
\begin{equation}
\frac{d T}{d S}=0,~~~~~~~~~~~~\frac{d^2 T}{d S^2}=0
\end{equation}
Solving the above mentioned equations we obtain the critical value of $S$ and $\tilde{\Phi}$ which are given as:
\begin{equation}
S_C=\beta~  ln \left(2 \left(\sqrt{3}-1\right)\right) ~~~~~~~~~~\tilde{\Phi}_C= \sqrt{1 - \pi~\tilde{\alpha}~(7+4 \sqrt{3}) } 
\label{eq27}
\end{equation}
By substituting the critical values from eqtn.\ref{eq27} in the expression for $T$ in eqtn.\ref{eq23} after having replaced $\tilde{Q}$ in terms of $\tilde{\Phi}$, the critical temperature, $ T_C$ is found to be $ T_C=11.2121\tilde{\alpha}$, and as we see it depends on the value of $\tilde{\alpha}$ which is quite different from the previous case (fixed $\tilde{Q}$ ensemble) where the critical temperature was a constant. Similar to what was previously done, using the value of the critical temperature we find the  relative temperature, $t$ that can be written as:
\begin{equation}
t = 
\frac{
\left( 2 \left( -1 + \sqrt{3} \right) \right)^{-1+s}
\left( -1 + 2 \left( -1 + \sqrt{3} \right) \right)^{3/2}
\left( \pi \tilde{\alpha} - \left( 7 + 4\sqrt{3} \right) \left( -1 + \left( 2 \left( -1 + \sqrt{3} \right) \right)^s \pi \tilde{\alpha} \rho^2 +e \right) \right)}{
\left( -1 + \left( 2 \left( -1 + \sqrt{3} \right) \right)^s \right)^{3/2}
\left( -2 \left( -1 + \sqrt{3} \right) + 8 \pi \tilde{\alpha} + 4\sqrt{3} \pi \tilde{\alpha} + 2 \left( -1 + \sqrt{3} \right) \left( 1 - 7 \pi \tilde{\alpha} - 4 \sqrt{3} \pi \tilde{\alpha} \right) \right)}
\end{equation}

where $e= \left( -1 + \left( 2 \left( -1 + \sqrt{3} \right) \right)^s \right) \left( -1 + \rho^2 \right)$ and
\begin{equation}
t=\frac{T}{T_C},~~~~s=\frac{S}{S_C},~~~~~\rho=\frac{\tilde{\Phi}}{\tilde{\Phi}_C}
\end{equation}
The rescaled Helmholtz free energy $f=F/F_C$ is obtained to be
\begin{equation}
f = \frac{\left(g-h+ k + 7 \left( 2 \left( -1 + \sqrt{3} \right)^s \left( -1 + \left( 2 \left( -1 + \sqrt{3} \right)^s \right) \right)s \rho^2
\log\left[ 2 \left( -1 + \sqrt{3} \right) \right] \right)
\right)}{\left( 
4 \sqrt{-3 + 2 \sqrt{3}} \left( -1 + \left( 2 \left( -1 + \sqrt{3} \right) \right)^s \right)^{3/2}
\pi \tilde{\alpha} \left( \sqrt{3} - 2 \log\left[ 2 \left( -1 + \sqrt{3} \right) \right] \right)
\right)}
\end{equation}

where, \begin{align*}
g = \Bigg( 
& 3 \left( 7 - 4 \sqrt{3} \right) \Bigg( 
\left( -1 + \left( 2 \left( -1 + \sqrt{3} \right) \right)^s \right)
\left( -1 + \rho^2 \right) 
\left( 2 - 2 \left( -2 + 2 \sqrt{3} \right)^s \right) \\
& \qquad\qquad\qquad\quad + \left( 2 \left( -1 + \sqrt{3} \right) \right)^s 
\log\left[ 2 \left( -1 + \sqrt{3} \right) \right] 
\Bigg) 
\Bigg)
\end{align*}.

 $h= \pi \tilde{\alpha}
\left(
2 - 2 \left( -2 + 2 \sqrt{3} \right)^s - 2 \left( 7 + 4 \sqrt{3} \right)
\left( -1 + \left( 2 \left( -1 + \sqrt{3} \right) \right)^s \right)^2
\rho^2 - 2 \left( -1 + \sqrt{3} \right)^s
\log\left[ 2 \left( -1 + \sqrt{3} \right) \right]
\right)$\\

$k=
-2^{2+s} \sqrt{3} \left( -1 + \sqrt{3} \right)^s + \sqrt{3} 4^{1+s} \left( -1 + \sqrt{3} \right)^{2s}$\\

\begin{figure}[ht]
\begin{center}
\includegraphics[width=.52\textwidth]{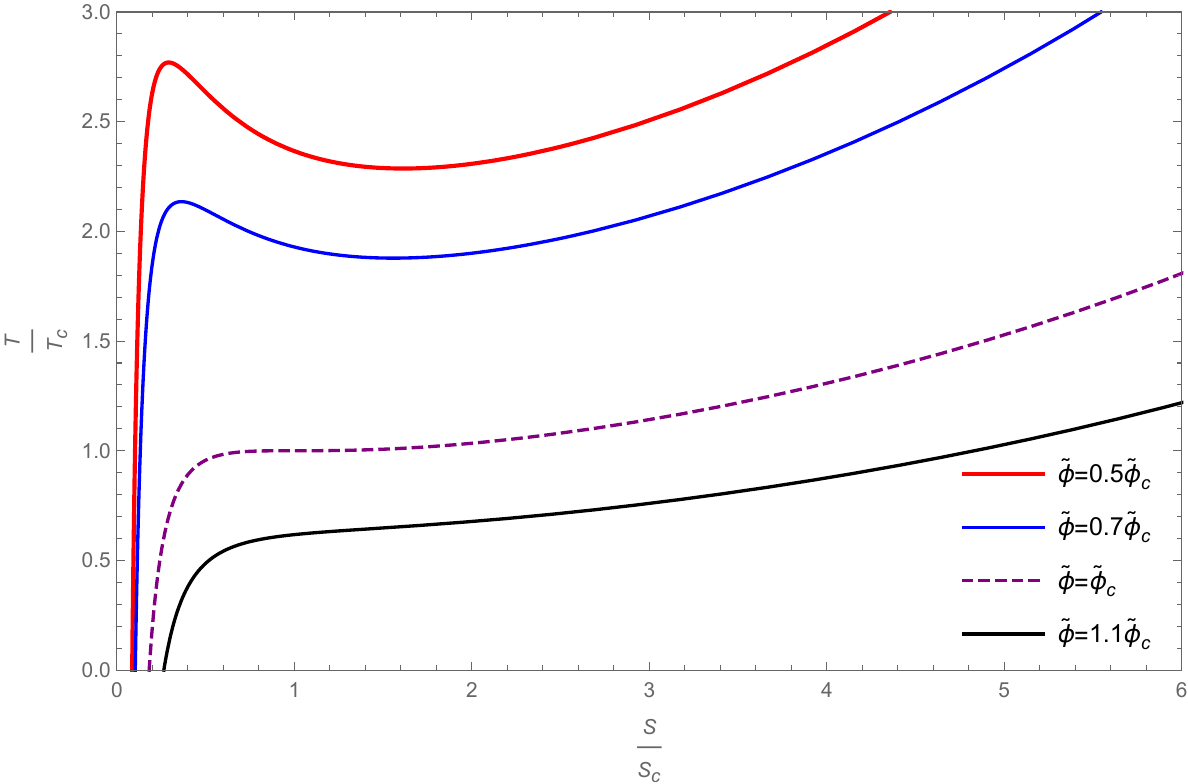}\vspace{3pt}
\includegraphics[width=.48\textwidth]{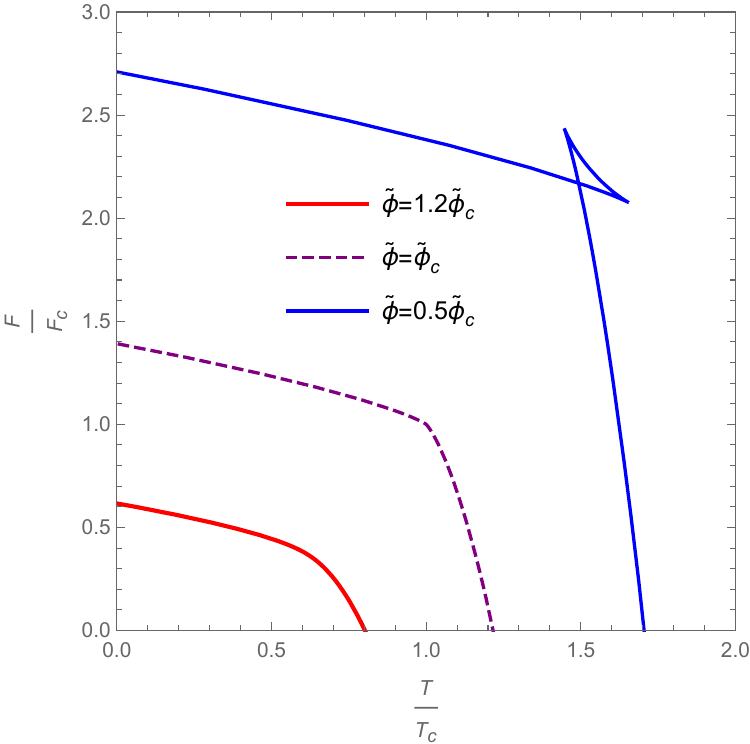}
\caption{$T-S$ and $F-T$ curves in the iso-voltage processes for $\tilde{\alpha}=0.01$ and $\beta=\frac{1}{\lambda}=83.33$}\label{fig3}
\end{center}
\end{figure}
 FIG. \ref{fig3} here illustrates both the $T-S$ and $F-T$ curves for the iso-voltage process for the Renyi modidfied flat 4D-EGB black hole. The plots quite distinctly show that below the critical value of $\tilde{\Phi}_C$ both the $T-S$ and $F-T$ curves show non-trivial behaviour where, the $T-S$ curve exhibits non-monotonic behaviour below $\tilde{\Phi}_C$ whereas the $F-T$ curve displays a swallowtail behaviour below the critical value of $\tilde{\Phi}_C$. Such behaviour is an indication for the possible Van der Waals-like first-order phase equilibrium in the iso-voltage processes when $0 < \tilde{\Phi} < \tilde{\Phi}_C$. However at the critical point $\tilde{\Phi} = \tilde{\Phi}_C$ we see that the second order phase transition is observed as represented by the dashed solid curve in both the figures.\\

It is important to highlight that first-order Van der Waals-like phase transitions do not typically arise in the thermodynamic behavior of non-AdS black holes. In the case of asymptotically flat charged black holes, the only known phase transitions are of the Davies type. However, our implementation of the RPST-like formalism reveals the emergence of a first-order Van der Waals-like phase transition even in the fixed $\tilde{\Phi}$ ensemble, akin to the behavior observed in the RPST framework for various charged AdS black holes.\\

\begin{figure}[ht]
\begin{center}
\includegraphics[width=.48\textwidth]{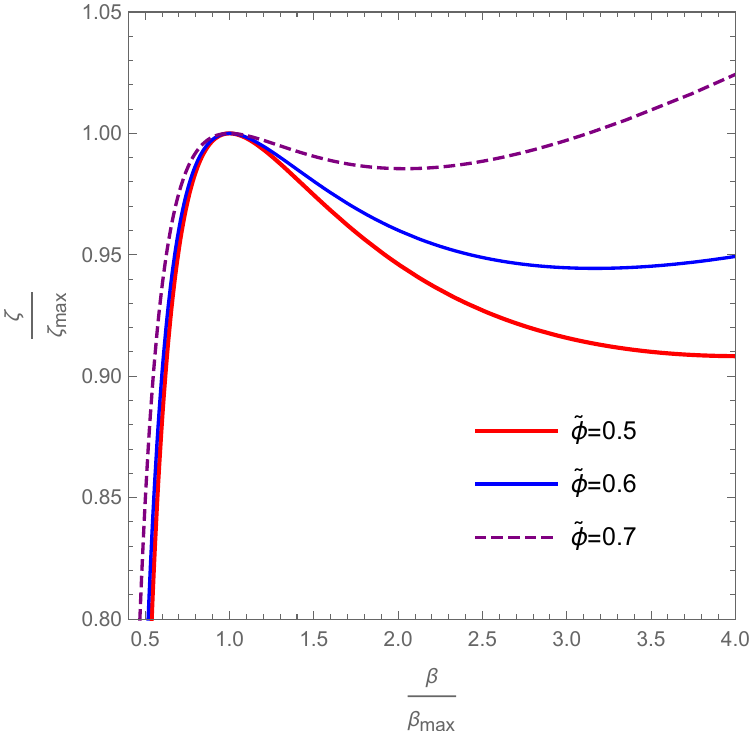}
\caption{$\zeta-\beta$ curves in the iso-voltage processes for $\tilde{\alpha}=0.002$ and $S=1$}\label{fig4}
\end{center}
\end{figure}

Finally, we analyze the $\zeta-\beta$ process in detail for the iso-voltage process. We first plot the expression for $\zeta$ as a function of $\beta$ using eqtn.(\ref{eq25}) after having replaced $\tilde{Q}$ in terms of $\tilde{\Phi}$ by the dint of eqtn.\ref{eq24}, keeping $\tilde{\Phi}$ and $S$ fixed. The corresponding plot is shown as  Fig.\ref{fig4}. The plot reveals that its behaviour is remarkably akin to the $\mu-C$ process in the AdS RPST formalism for the non-modified EGB black hole.

\section{Thermodynamic Geometry and Thermodynamic Topology of the Rényi modified Flat 4D-EGB Black Hole.}
\label{sec:GTD}

\subsection{Thermodynamic Geometry.}

For the flat 4D-EGB black hole, which arises in a regularized version of higher-dimensional Gauss-Bonnet gravity scaled down to four dimensions, the GTD metric encapsulates the thermodynamic phase space, where the coordinates include extensive variables like entropy, $S$ and electric charge, $\tilde{Q}$ and intensive variables like temperature, $T$ and Gauss-Bonnet(GB) coupling, $\tilde{\alpha}$. The GTD formalism constructs a Riemannian metric on this phase space, enabling the analysis of thermodynamic interactions and phase transitions through geometric properties, such as curvature singularities, which correspond to critical points or phase transitions in the black hole’s thermodynamics. For the flat 4D-EGB black hole, the GTD metric reveals insights into its stability and critical behaviour, particularly in the presence of the Gauss-Bonnet coupling, and investigates if certain modifications arise in the black hole’s thermodynamic properties compared to standard Einstein gravity, and therefore offering a geometric interpretation of its entropy, temperature, and any possible potential instabilities.\\

\underline{{\textbf{Fixed $(\tilde{Q})$ ensemble}:}}\\

We write the GTD metric for the flat 4D-EGB black hole in the fixed $(\tilde{Q})$ ensemble from the general metric given in  eqtn.\ref{eq12}. We consider the thermodynamic potential $\varphi$ to be the mass $\tilde{M}$ for the Renyi modified 4D-EGB black hole in  the fixed $(\tilde{Q})$ ensemble as obtained in equation eqtn.\ref{eq22}. The GTD metric is thereby given as :- 
$$g  =S \left(\frac{\partial \tilde{M}}{\partial S}\right)\left(- \frac{\partial^2 \tilde{M}}{\partial S^2} dS^2 + \frac{\partial^2 \tilde{M}}{\partial \tilde{Q}^2} d\tilde{Q}^2 \right) $$  
\\

From the above metric we calculate the GTD scalar for  the Renyi modified flat black hole in the fixed $(\tilde{Q})$ ensemble. Although we do not present here the explicit derivation or the full expressions for the GTD scalar curvature due to their substantial length, we note that these can be obtained through straightforward and routine mathematical manipulations. In what follows, we provide a detailed analysis of the GTD thermodynamic geometry corresponding to the Rényi-modified flat 4D Einstein-Gauss-Bonnet (EGB) black hole within the fixed ensemble characterized by constant \(\tilde{Q}\)\\

\begin{figure}[h]	
	\centering
	\begin{subfigure}{0.40\textwidth}
		\includegraphics[width=\linewidth]{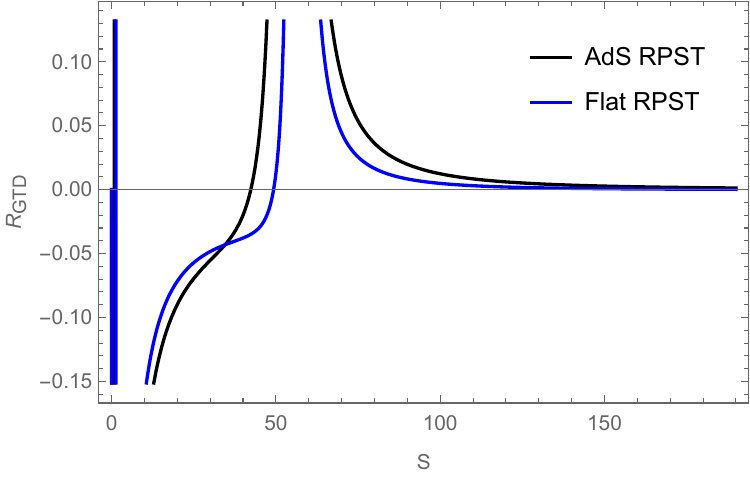}
		\caption{Fixed $(\tilde{Q})$ ensemble for $\tilde{\alpha}=0.002$ and $\tilde{Q}=0.4$.}
		\label{5a}
		\end{subfigure}
		\hspace{0.5cm}
		\begin{subfigure}{0.40\textwidth}
		\includegraphics[width=\linewidth]{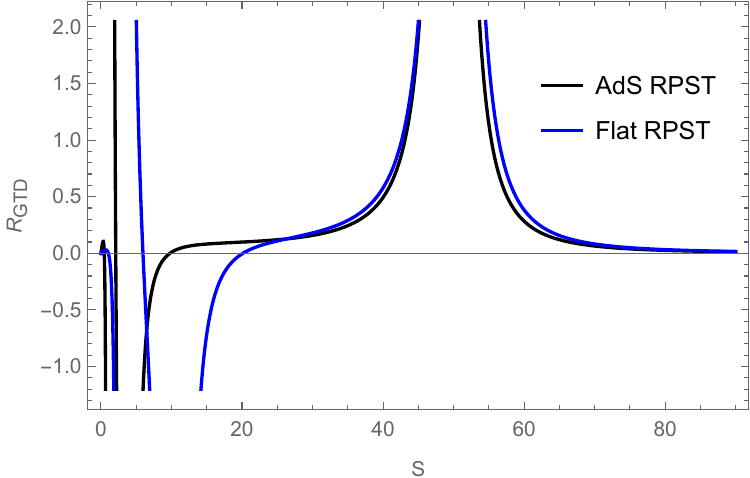}
		\caption{Fixed $(\tilde{\Phi})$ ensemble for $\tilde{\alpha}= 0.01$ and $\tilde{\Phi}=0.4$.}
		\label{5b}
		\end{subfigure}
		\caption{The GTD scalar versus entropy plot for the RPST like formalism of the Renyi modified flat black hole(Blue) and the RPST formalism of the AdS black hole(Black). }
	\label{5}
 \end{figure}

We draw a plot for the GTD scalar, $R_{GTD}$ versus entropy for 
$\tilde{\alpha}= 0.002$ and $\tilde{Q}=0.4$ for the Renyi modified flat 4D-EGB black hole along with the AdS 4D-EGB black hole as can be seen from Fig.\ref{5}. And from Fig.\ref{5a} we find that the thermodynamic geometry of the Renyi modified RPST like formalism for flat 4D-EGB black hole is \textbf{remarkably similar} to that of the RPST formalism of the AdS 4D-EGB black hole in the fixed $(\tilde{Q})$ ensemble, where we see a direct relationship between the inverse Renyi parameter $\beta$ in the flat black hole with that of the central charge $C$ in the AdS black hole. We see that for $C=53$ and $\beta=\frac{1}{\lambda}=83.33$ the thermodynamic geometry of the both the black holes coincide for the fixed $(\tilde{Q})$ ensemble as can be seen from the plot itself. It is important here to mention that the plot for thermodynamic scalar reveals the fact that the central charge, $C$ and the Renyi parameter, $\lambda$ are inversely proportional to each other i.e. a small Renyi parameter, $\lambda$ is equivalent to a large central charge, $C$ and vice-versa.\\

\underline{{\textbf{Fixed $(\tilde{\Phi})$ ensemble}:}}\\

We write the GTD metric for the flat 4D-EGB black hole in the fixed $(\tilde{\Phi})$ ensemble from the general metric given in  eqtn.\ref{eq12}. We consider the thermodynamic potential $\varphi$ to be the mass $\tilde{M}$ for the Renyi modified 4D-EGB black hole in  the fixed $(\tilde{\Phi})$ ensemble as obtained in equation eqtn.\ref{eq22} after having replaced $\tilde{Q}$ in terms of $\tilde{\Phi}$ by the dint of eqtn.\ref{eq24}. The GTD metric is thereby given as :- 
$$g  =S \left(\frac{\partial \tilde{M}}{\partial S}\right)\left(- \frac{\partial^2 \tilde{M}}{\partial S^2} dS^2 + \frac{\partial^2 \tilde{M}}{\partial \tilde{\Phi}^2} d\tilde{\Phi}^2 \right) $$  
\\

From the above metric we calculate the GTD scalar for  the Renyi modified flat black hole in the fixed $(\tilde{\Phi})$ ensemble. Although we do not present here the explicit derivation or the full expressions for the GTD scalar curvature due to their substantial length, we note that these can be obtained through straightforward and routine mathematical manipulations. In what follows, we provide a detailed analysis of the GTD thermodynamic geometry corresponding to the Rényi-modified flat 4D Einstein-Gauss-Bonnet (EGB) black hole within the fixed ensemble characterized by constant \(\tilde{\Phi}\)\\

We draw a plot for the GTD scalar, $R_{GTD}$ versus entropy for $\tilde{\alpha}= 0.01$ and $\tilde{\Phi}=0.4$ for  the Renyi modified flat 4D-EGB black hole along with the AdS 4D-EGB black hole as can be seen from Fig.\ref{5}. And from Fig.\ref{5b} we find that the thermodynamic geometry of the Renyi modified RPST like formalism for flat 4D-EGB black hole is \textbf{remarkably similar} to that of the RPST formalism of the AdS 4D-EGB black hole in the fixed $(\tilde{\Phi})$ ensemble, where we see a direct relationship between the inverse Renyi parameter $\beta$ in the flat black hole with that of the central charge $C$ in the AdS black hole. We see that for $C=44$ and $\beta=\frac{1}{\lambda}=83$ the thermodynamic geometry of the both the black holes coincide for the fixed $(\tilde{\Phi})$ ensemble as can be seen from the plot itself. It is important here to mention that just like the previous case for fixed $(\tilde{Q})$ ensemble, the plot here for thermodynamic scalar also shows that the central charge, $C$ and the Renyi parameter, $\lambda$ are inversely proportional to each other i.e. a small Renyi parameter, $\lambda$ is equivalent to a large central charge, $C$ and vice-versa.  

 \subsection{Thermodynamic Topology.}

The topological approach as was previously discussed in the introduction, provides a robust framework to analyse black hole thermodynamics by regarding black hole solutions as topological defects in their thermodynamic phase space. This method systematically proceeds as follows:\\

 The procedure begins with the construction of the off-shell free energy which is already mentioned in eq. (\ref{eqF}),
 $$\mathcal{F}(S;\tau) = M(S) - \frac{S}{\tau},$$
where $M(S)$ is the black hole mass as a function of entropy and $\tau$ is the inverse temperature of a thermal bath. We then perform the differentiation of the off-shell free energy with respect to entropy which further gives
\begin{equation}
    \phi^S(S) = \frac{\partial \mathcal{F}(S;\tau)}{\partial S} = T_H(S) - \frac{1}{\tau},
\end{equation}
where $T_H(S)$ is the Hawking temperature. By putting $\phi^S(S) = 0$ one can find $\tau$ as a function of $S$ and then draw a parametric plot in the ($\tau$, $S$ ) plane to locate the zeros of the vector field which in turn also identifies the number of phase transitions in the thermodynamic phase space of the black hole. To further analyse these points, one introduces a two-dimensional vector field
\begin{equation}
    \vec{\phi}(S,{\Theta}) = \big(\phi^S,\;\phi^{\Theta}\big), 
    \qquad \phi^{\Theta} = -\cot{\Theta} \csc{\Theta},
\end{equation}
where ${\Theta} \in (0,\pi)$ is an auxiliary coordinate. Vector plots in the $(S,{\Theta})$ plane illustrate the circulation of the field around the zeros, thereby revealing their local topological structure. To quantify this structure, one traces a small closed loop around each zero, parametrized as
\begin{equation}
    S(\theta) = S_0 + \rho_S\cos\theta, \qquad 
    \Theta(\theta) = \tfrac{\pi}{2} + \rho_\Theta\sin\theta, \qquad \theta\in[0,2\pi],
\end{equation}
Also, $\rho_S$ and $\rho_\Theta$ are parameters that define the dimensions of the contour to be formed and $S_0$ represents the center point around which the contour is created. The winding numbers  quantify how the vector field deflects along contours enclosing each zero point. The mathematical relation between deflection angle $\Omega$ and the vector field is given by the contour integration,
\begin{equation}
 \Omega(\theta) = \int_{0}^{\theta} \epsilon_{12} \ n^{1} \ \partial_{\theta} n^{2} \, d\theta
\end{equation}
where, $n^{1}$ and $n^{2}$ can be found through eq. (\ref{eqn}). The winding number associated with the defect is then obtained as
\begin{equation}
    w = \frac{1}{2\pi}\,\Delta \Omega(\theta)\Big|_{0}^{2\pi},
\end{equation}
where $\Delta \Omega$ denotes the total change in phase angle along the loop. A net increase of $+2\pi$ gives $w=+1$, while a decrease of $-2\pi$ gives $w=-1$, and if the phase returns to its initial value, the winding number is zero. The sum of all winding numbers defines the total topological charge,
\begin{equation}
    W = \sum_i w_i,
\end{equation}
which is a topological invariant of the thermodynamic system.\\

\underline{{\textbf{Fixed $(\tilde{Q})$ ensemble for the Rényi  modified flat black hole}:}}\\
 
\begin{figure}[h]	
	\centering
	\begin{subfigure}{0.3\textwidth}
		\includegraphics[width=\linewidth]{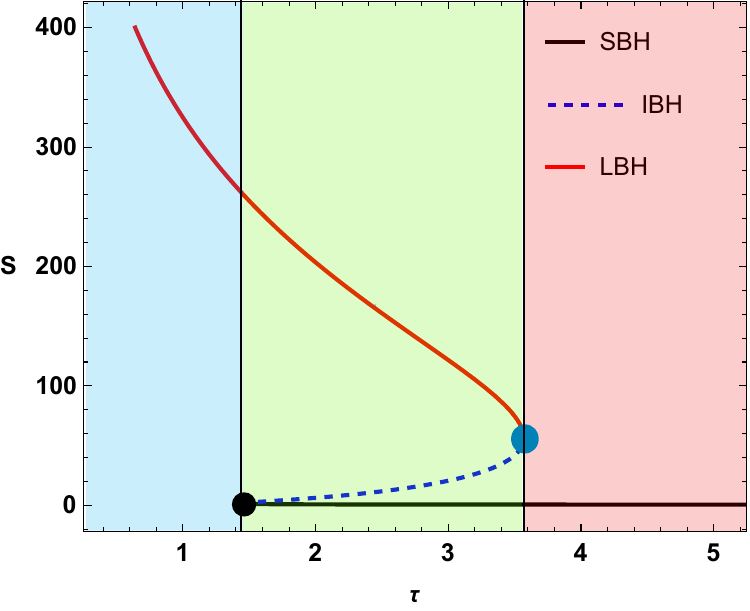}
		\caption{$S$-$\tau$ plot in fixed ($\tilde{Q}$) ensemble.}
		\label{6a}
		\end{subfigure}
		\begin{subfigure}{0.3\textwidth}
		\includegraphics[width=\linewidth]{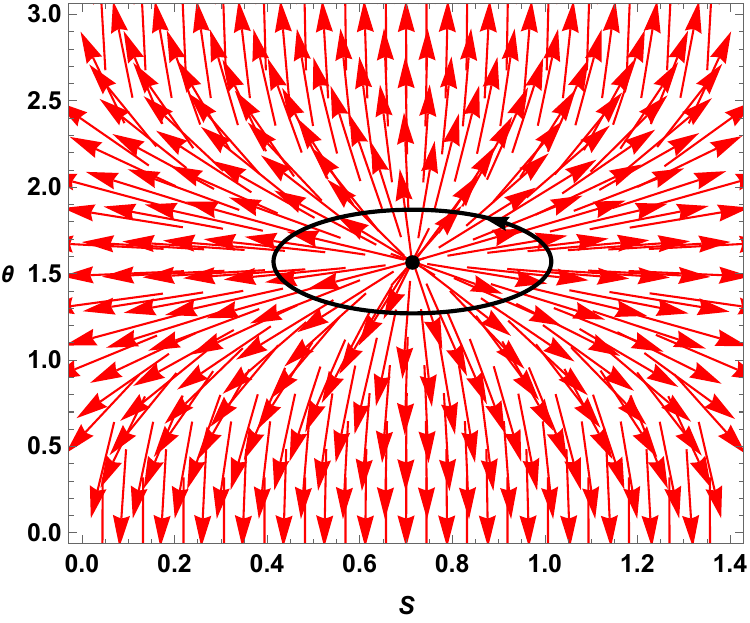}
		\caption{ Vector plot around $S=0.713$.}
		\label{6b}
		\end{subfigure}
		\begin{subfigure}{0.3\textwidth}
		\includegraphics[width=\linewidth]{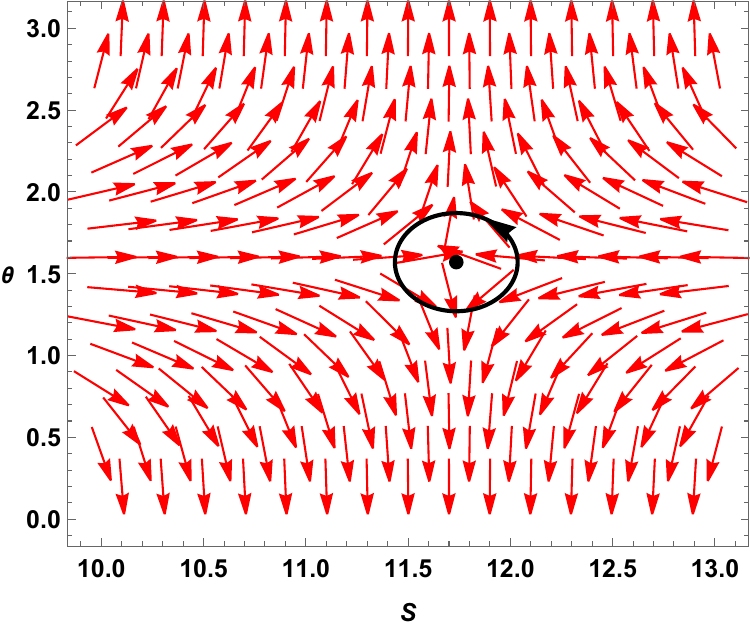}
		\caption{ Vector plot around $S=11.736$.}
		\label{6c}
		\end{subfigure}
		
		\begin{subfigure}{0.3\textwidth}
		\includegraphics[width=\linewidth]{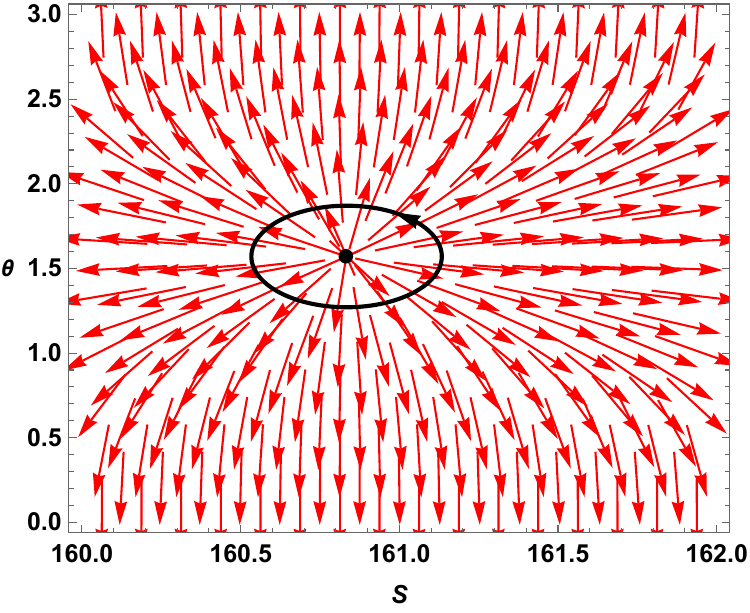}
		\caption{ Vector plot around $S=160.834$.}
		\label{6d}
		\end{subfigure}
		\begin{subfigure}{0.3\textwidth}
		\includegraphics[width=\linewidth]{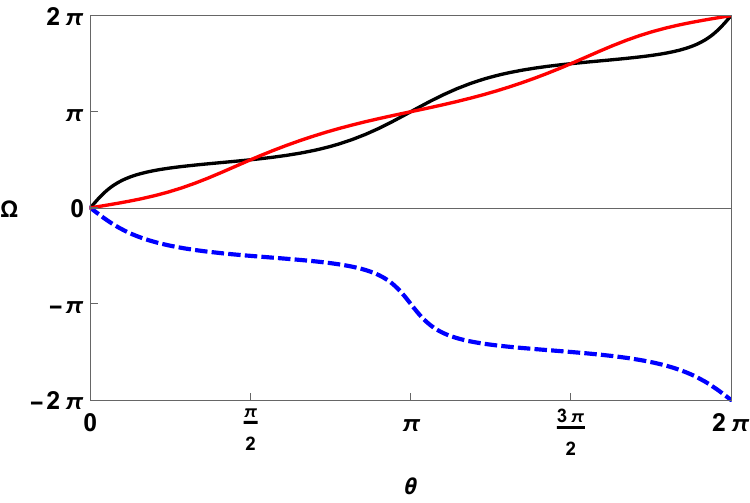}
		\caption{ Contour plot.}
		\label{6e}
		\end{subfigure}
		\caption{Thermodynamic topology for the Rényi modified flat 4D-EGB black hole in fixed ($\tilde{Q}$) ensemble for $\tilde{\alpha}= 0.002$ and $\tilde{Q}=0.4$. }
	\label{6}
 \end{figure}

For $\tilde{Q} = 0.4$, $\tilde{\alpha} = 0.002$, $\beta = 83.33$, we plot the $\tau$ vs $S$ curve for the  Rényi  modified flat 4D-EGB black hole in the fixed $\tilde{Q}$-ensemble as is shown in Fig.~\ref{6a}. This plot reveals three black hole branches: a small black hole branch (solid black curve); an intermediate black hole branch (blue dotted curve); and a large black hole branch (red solid curve). Now to calculate the topological charge of the black hole we choose a random value of $\tau$, say $\tau = 2.5$, and calculate the zero points of the vector field. On the figures Fig.~\ref{6b}, Fig.~\ref{6c} and Fig.~\ref{6d}, the vector field $(n^{1}, n^{2})$ is plotted in the $(S, \tau)$ plane, where the zero points for $\tau = 2.5$ are observed at $S = 0.713$, $S = 11.736$ and $S = 160.834$ as can be seen from the respective plots. The topological charge is finally computed from Fig.~\ref{6e}, where the winding number corresponding to $S = 0.713$ and $S = 160.834$ is $+1$ as represented by the black and red solid curves respectively, whereas the winding number for $S = 11.736$ comes out to be $-1$ as is seen from the blue dashed curve in the plot. Therefore, winding numbers for the small and large black hole(BH) branch comes out to be $+1$ whereas for the intermediate BH branch the winding number turns out to be $-1$. It is important, however, to note that positive winding numbers represent stable BH branches whereas negative winding numbers represent the unstable BH branch. It can therefore be concluded that the small and large BH branches here are stable whereas the intermediate BH branch turns out to be unstable. The total topological charge is therefore obtained by adding all the winding numbers as given by: 
$$W = +1 + 1 - 1 = +1$$ .

\underline{{\textbf{Fixed $(\tilde{Q})$ ensemble for the AdS RPST black hole}:}}\\
 
\begin{figure}[h]	
	\centering
	\begin{subfigure}{0.3\textwidth}
		\includegraphics[width=\linewidth]{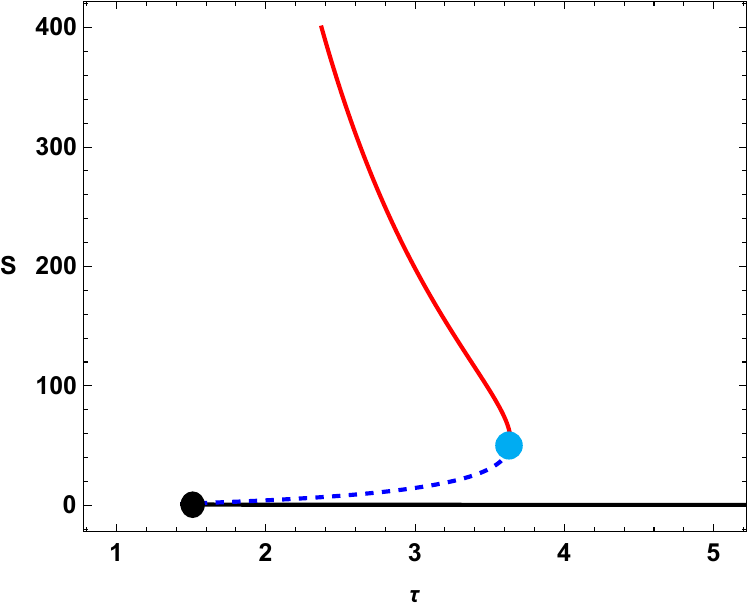}
		\caption{$S$-$\tau$ plot in fixed ($\tilde{Q}$) ensemble.}
		\label{7a}
		\end{subfigure}
		\begin{subfigure}{0.3\textwidth}
		\includegraphics[width=\linewidth]{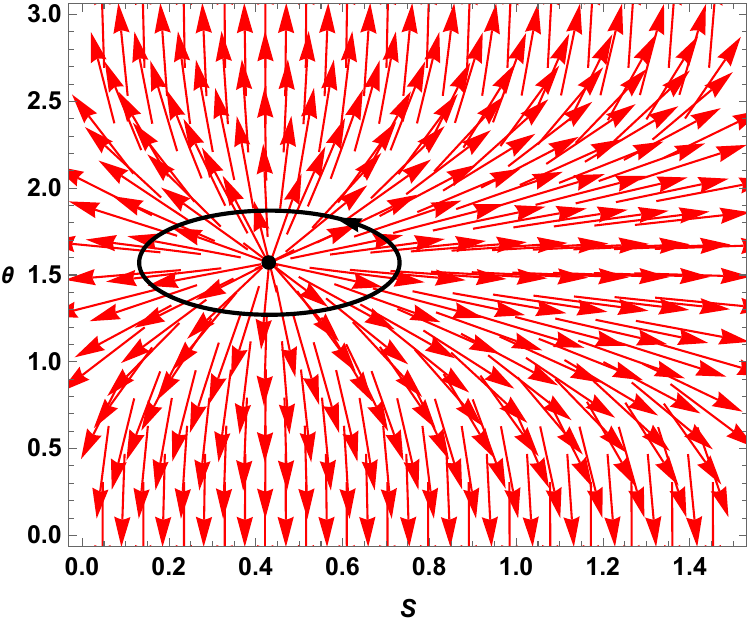}
		\caption{ Vector plot around $S=0.431$.}
		\label{7b}
		\end{subfigure}
		\begin{subfigure}{0.3\textwidth}
		\includegraphics[width=\linewidth]{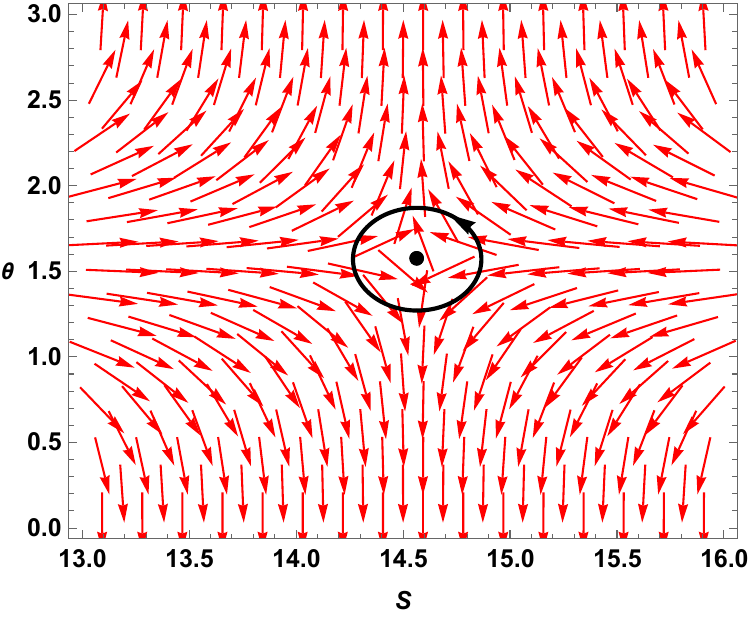}
		\caption{ Vector plot around $S=14.566$.}
		\label{7c}
		\end{subfigure}
		
		\begin{subfigure}{0.3\textwidth}
		\includegraphics[width=\linewidth]{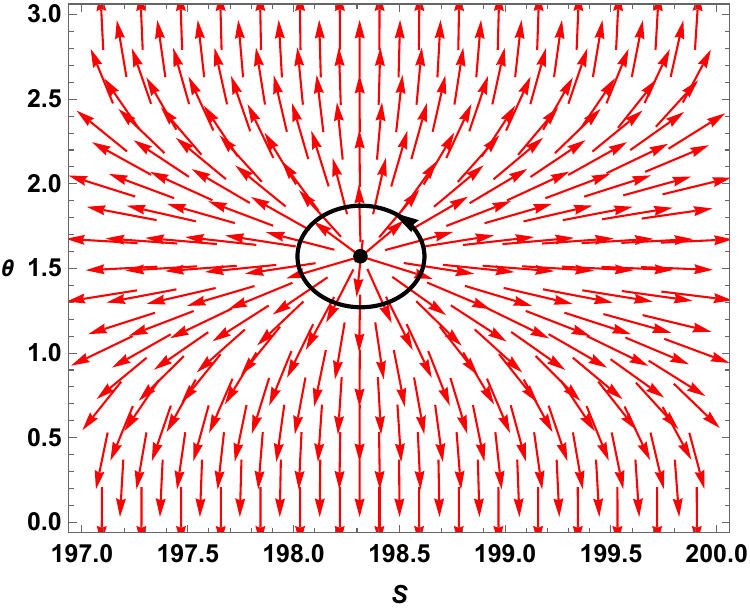}
		\caption{ Vector plot around $S=198.319$.}
		\label{7d}
		\end{subfigure}
		\begin{subfigure}{0.3\textwidth}
		\includegraphics[width=\linewidth]{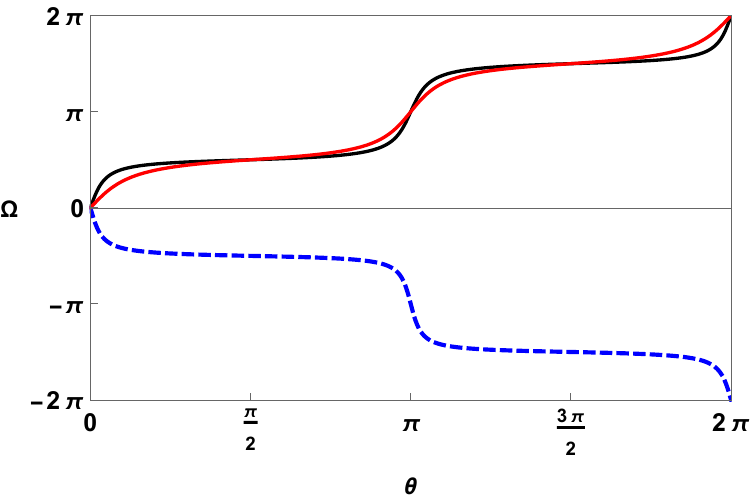}
		\caption{ Contour plot.}
		\label{7e}
		\end{subfigure}
		\caption{Thermodynamic topology for the AdS RPST 4D-EGB black hole in fixed ($\tilde{Q}$) ensemble for $\tilde{\alpha}= 0.002$ and $\tilde{Q}=0.4$. }
	\label{7}
 \end{figure}

For $\tilde{Q} = 0.4$, $\tilde{\alpha} = 0.002$, $C = 53$, we plot the $\tau$ vs $S$ curve for the  AdS RPST 4D-EGB black hole in the fixed $\tilde{Q}$-ensemble as is shown in Fig.~\ref{7a}. This plot reveals three black hole branches: a small black hole branch (solid black curve); an intermediate black hole branch (blue dotted curve); and a large black hole branch (red solid curve). Now to calculate the topological charge of the black hole we choose a random value of $\tau$, 
say $\tau = 3$, and calculate the zero points of the vector field. On the figures Fig.~\ref{7b}, Fig.~\ref{7c} and Fig.~\ref{7d}, the vector field $(n^{1}, n^{2})$ is plotted in the $(S, \tau)$ plane, where the zero points for $\tau = 3$ are observed at $S = 0.431$, $S = 14.566$ and $S = 198.319$ as can be seen from the respective plots. The topological charge is finally computed from Fig.~\ref{7e}, where the winding number corresponding to $S = 0.431$ and $S = 198.319$ is $+1$ as represented by the black and red solid curves respectively, whereas the winding number for $S = 14.566$ comes out to be $-1$ as is seen from the blue dashed curve in the plot. Therefore, winding numbers for the small and large black hole(BH) branch comes out to be $+1$ whereas for the intermediate BH branch the winding number turns out to be $-1$. It is important, however, to note that positive winding numbers represent stable BH branches whereas negative winding numbers represent the unstable BH branch. It can therefore be concluded that the small and large BH branches here are stable whereas the intermediate BH branch turns out to be unstable. The total topological charge is therefore obtained by adding all the winding numbers as given by: 
$$W = +1 + 1 - 1 = +1$$ .

\underline{{\textbf{Fixed $(\tilde{\Phi})$ ensemble for the Rényi  modified flat black hole}:}}\\
 
\begin{figure}[h]	
	\centering
	\begin{subfigure}{0.3\textwidth}
		\includegraphics[width=\linewidth]{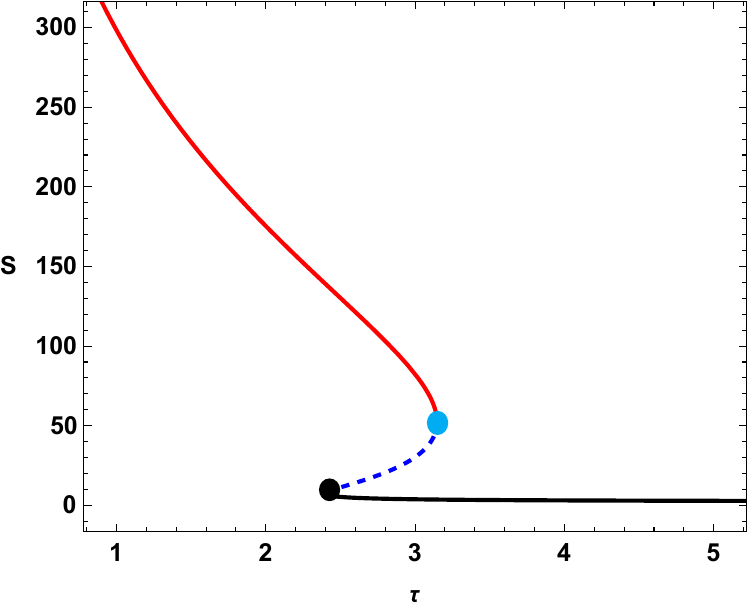}
		\caption{$S$-$\tau$ plot in fixed ($\tilde{\Phi}$) ensemble.}
		\label{8a}
		\end{subfigure}
		\begin{subfigure}{0.3\textwidth}
		\includegraphics[width=\linewidth]{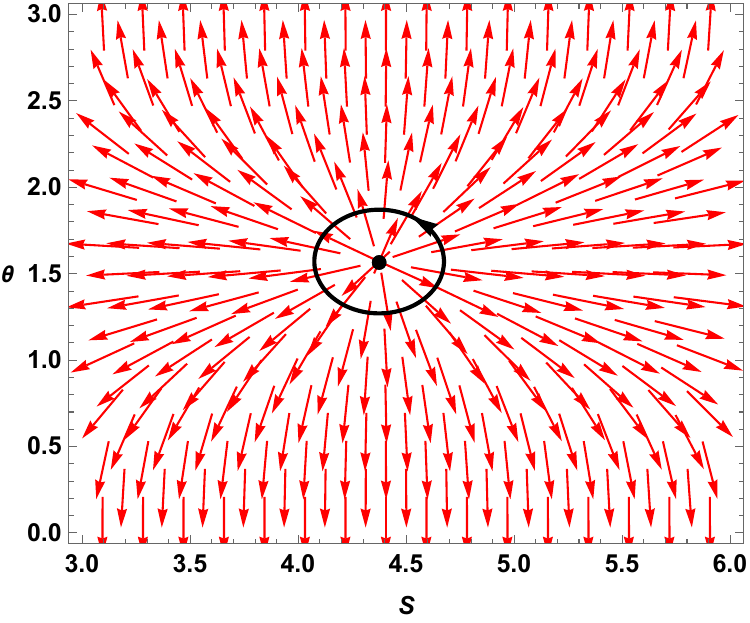}
		\caption{ Vector plot around $S=4.374$.}
		\label{8b}
		\end{subfigure}
		\begin{subfigure}{0.3\textwidth}
		\includegraphics[width=\linewidth]{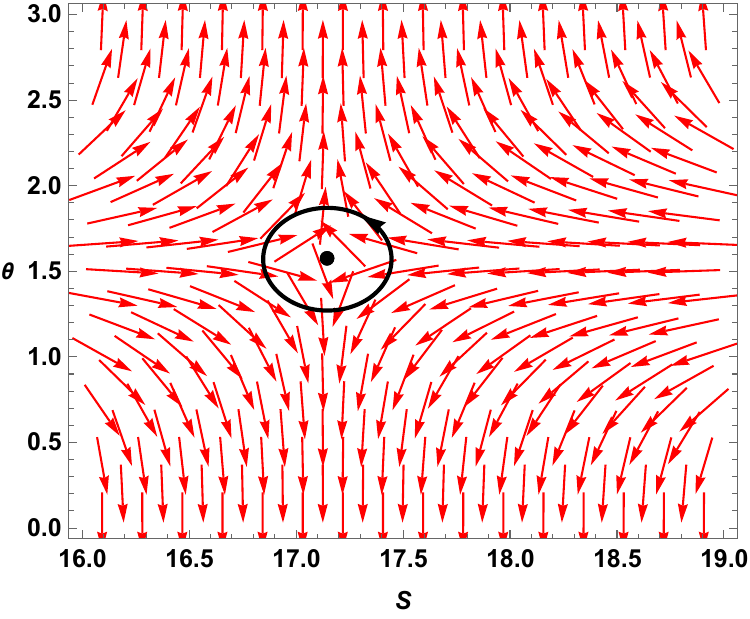}
		\caption{ Vector plot around $S=17.146$.}
		\label{8c}
		\end{subfigure}
		
		\begin{subfigure}{0.3\textwidth}
		\includegraphics[width=\linewidth]{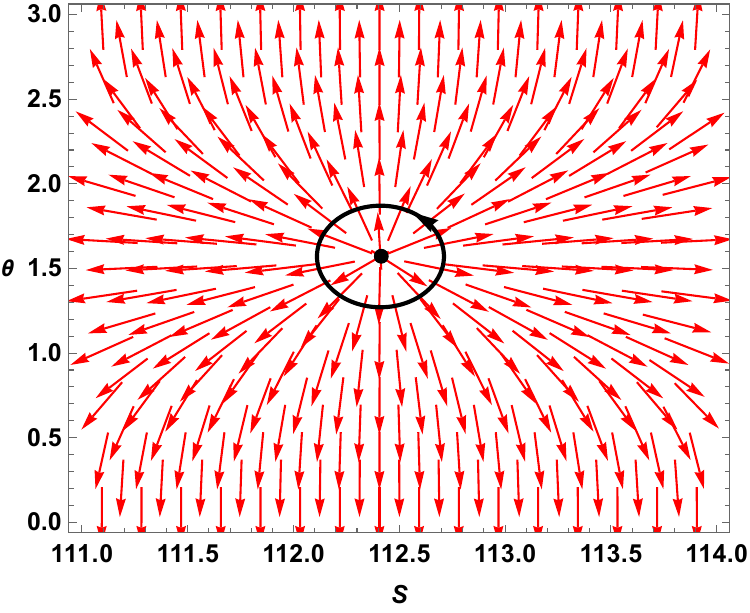}
		\caption{ Vector plot around $S=112.411$.}
		\label{8d}
		\end{subfigure}
		\begin{subfigure}{0.3\textwidth}
		\includegraphics[width=\linewidth]{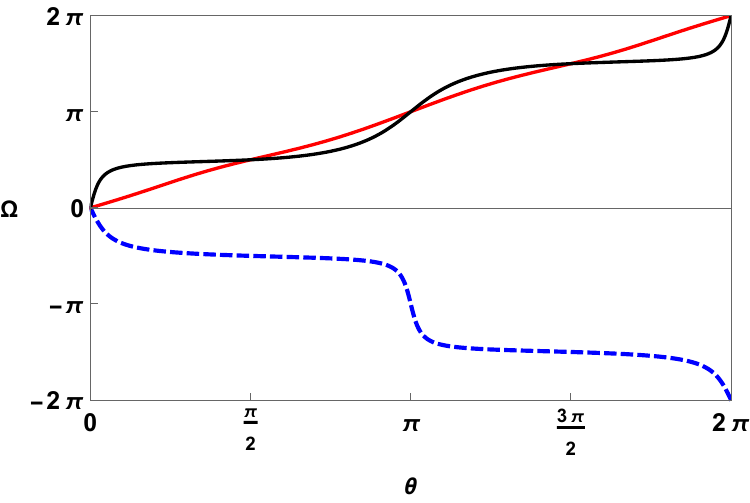}
		\caption{ Contour plot.}
		\label{8e}
		\end{subfigure}
		\caption{Thermodynamic topology for the Rényi modified flat 4D-EGB black hole in fixed ($\tilde{\Phi}$) ensemble for $\tilde{\alpha}= 0.01$ and $\tilde{\Phi}=0.4$. }
	\label{8}
 \end{figure}

For $\tilde{\Phi} = 0.4$, $\tilde{\alpha} = 0.01$, $\beta = 83$, we plot the $\tau$ vs $S$ curve for the  Rényi  modified flat 4D-EGB black hole in the fixed $\tilde{\Phi}$-ensemble as is shown in Fig.~\ref{8a}. This plot reveals three black hole branches: a small black hole branch (solid black curve); an intermediate black hole branch (blue dotted curve); and a large black hole branch (red solid curve). Now to calculate the topological charge of the black hole we choose a random value of $\tau$, say $\tau = 2.7$, and calculate the zero points of the vector field. On the figures Fig.~\ref{8b}, Fig.~\ref{8c} and Fig.~\ref{8d}, the vector field $(n^{1}, n^{2})$ is plotted in the $(S, \tau)$ plane, where the zero points for $\tau = 2.7$ are observed at $S = 4.374$, $S = 17.146$ and $S = 112.411$ as can be seen from the respective plots. The topological charge is finally computed from Fig.~\ref{8e}, where the winding number corresponding to $S = 4.374$ and $S = 112.411$ is $+1$ as represented by the black and red solid curves respectively, whereas the winding number for $S = 17.146$ comes out to be $-1$ as is seen from the blue dashed curve in the plot. Therefore, winding numbers for the small and large black hole(BH) branch comes out to be $+1$ whereas for the intermediate BH branch the winding number turns out to be $-1$. It is important, however, to note that positive winding numbers represent stable BH branches whereas negative winding numbers represent the unstable BH branch. It can therefore be concluded that the small and large BH branches here are stable whereas the intermediate BH branch turns out to be unstable. The total topological charge is therefore obtained by adding all the winding numbers as given by: 
$$W = +1 + 1 - 1 = +1$$ .

 \underline{{\textbf{Fixed $(\tilde{\Phi})$ ensemble for the AdS RPST black hole}:}}\\
 
\begin{figure}[h]	
	\centering
	\begin{subfigure}{0.3\textwidth}
		\includegraphics[width=\linewidth]{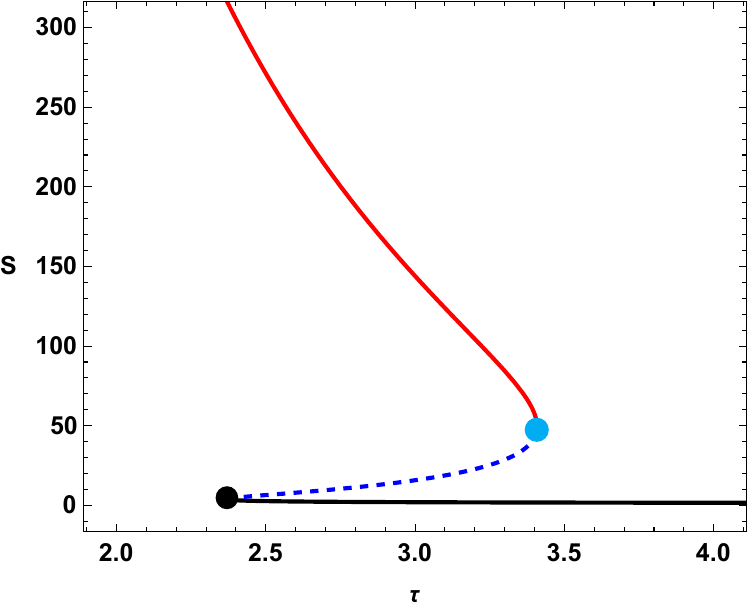}
		\caption{$S$-$\tau$ plot in fixed ($\tilde{\Phi}$) ensemble.}
		\label{9a}
		\end{subfigure}
		\begin{subfigure}{0.3\textwidth}
		\includegraphics[width=\linewidth]{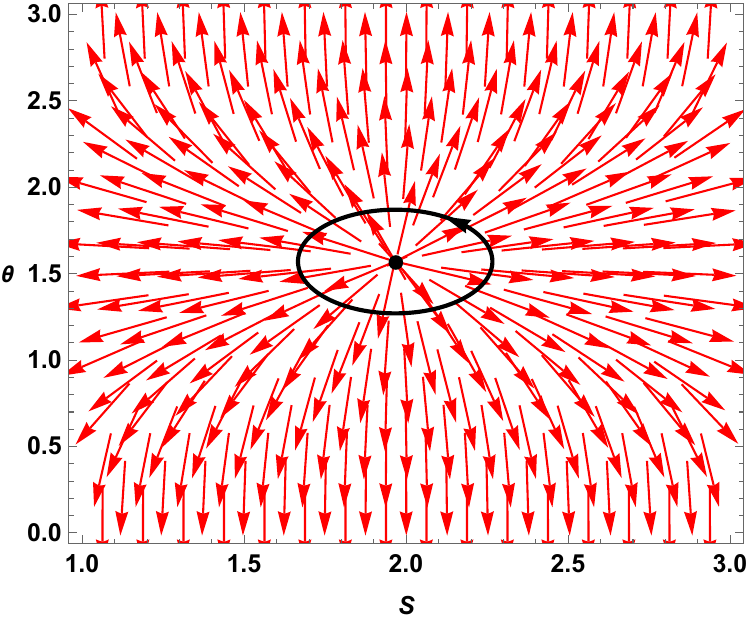}
		\caption{ Vector plot around $S=1.966$.}
		\label{9b}
		\end{subfigure}
		\begin{subfigure}{0.3\textwidth}
		\includegraphics[width=\linewidth]{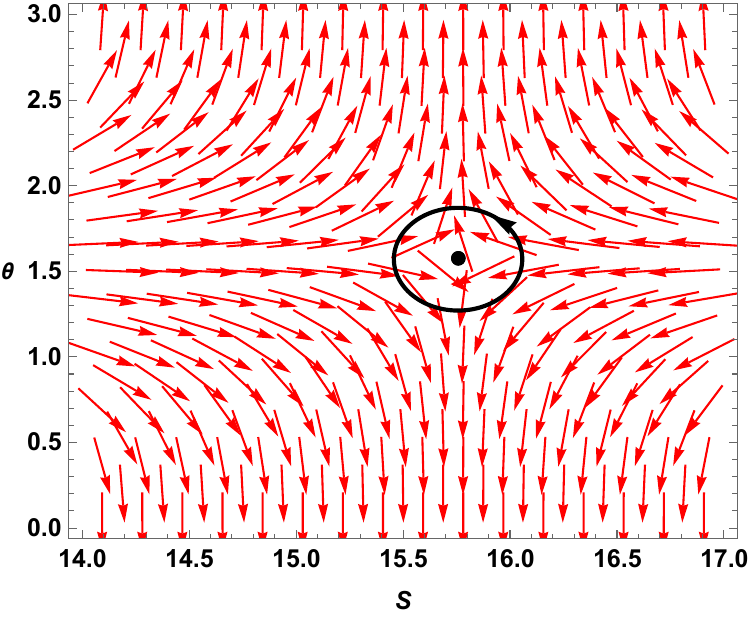}
		\caption{ Vector plot around $S=15.757$.}
		\label{9c}
		\end{subfigure}
		
		\begin{subfigure}{0.3\textwidth}
		\includegraphics[width=\linewidth]{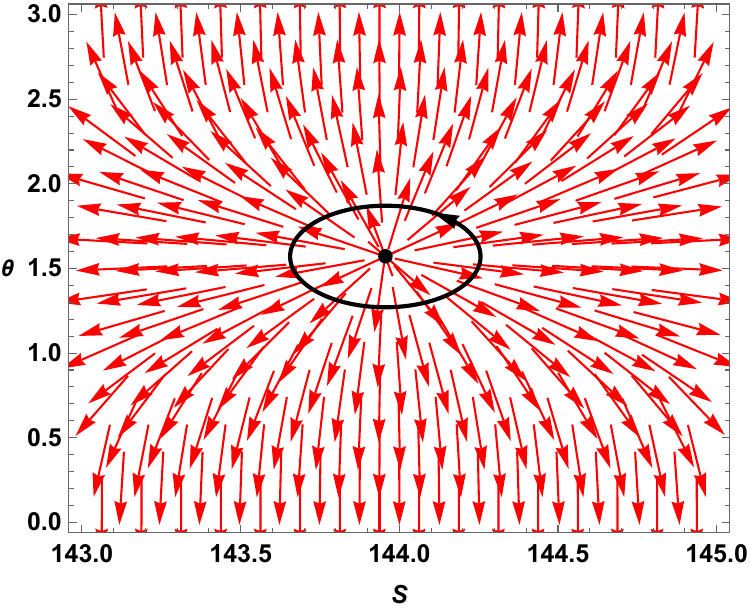}
		\caption{ Vector plot around $S=143.956$.}
		\label{9d}
		\end{subfigure}
		\begin{subfigure}{0.3\textwidth}
		\includegraphics[width=\linewidth]{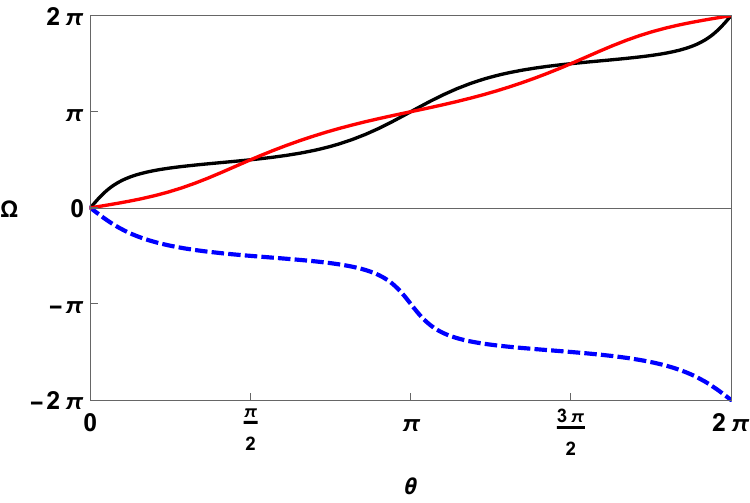}
		\caption{ Contour plot.}
		\label{9e}
		\end{subfigure}
		\caption{Thermodynamic topology for the AdS RPST 4D-EGB black hole in fixed ($\tilde{\Phi}$) ensemble for $\tilde{\alpha}= 0.01$ and $\tilde{\Phi}=0.4$. }
	\label{9}
 \end{figure}

For $\tilde{\Phi} = 0.4$, $\tilde{\alpha} = 0.01$, $C = 44$, we plot the $\tau$ vs $S$ curve for the  Rényi  modified flat 4D-EGB black hole in the fixed $\tilde{\Phi}$-ensemble as is shown in Fig.~\ref{9a}. This plot reveals three black hole branches: a small black hole branch (solid black curve); an intermediate black hole branch (blue dotted curve); and a large black hole branch (red solid curve). Now to calculate the topological charge of the black hole we choose a random value of $\tau$, say $\tau = 3$, and calculate the zero points of the vector field. On the figures Fig.~\ref{9b}, Fig.~\ref{9c} and Fig.~\ref{9d}, the vector field $(n^{1}, n^{2})$ is plotted in the $(S, \tau)$ plane, where the zero points for $\tau = 3$ are observed at $S = 1.966$, $S = 15.757$ and $S = 143.956$ as can be seen from the respective plots. The topological charge is finally computed from Fig.~\ref{9e}, where the winding number corresponding to $S = 1.966$ and $S = 143.956$ is $+1$ as represented by the black and red solid curves respectively, whereas the winding number for $S = 15.757$ comes out to be $-1$ as is seen from the blue dashed curve in the plot. Therefore, winding numbers for the small and large black hole(BH) branch comes out to be $+1$ whereas for the intermediate BH branch the winding number turns out to be $-1$. It is important, however, to note that positive winding numbers represent stable BH branches whereas negative winding numbers represent the unstable BH branch. It can therefore be concluded that the small and large BH branches here are stable whereas the intermediate BH branch turns out to be unstable. The total topological charge is therefore obtained by adding all the winding numbers as given by: 
$$W = +1 + 1 - 1 = +1$$ .

We can therefore conclude from the topological analysis of both the black hole systems that both AdS RPST and Renyi modified flat 4D-EGB black hole belong to the same topological class i.e. W=$+1$ for both the fixed ($\tilde{Q}$) and ($\tilde{\Phi}$) ensembles. This finding is in complete concordance with the thermodynamic geometric analysis done in the previous subsection where it revealed that both the black hole systems are remarkably alike with respect to their thermodynamic characteristics.  
\newpage
\section{Conclusions:}
\label{sec:conclusion}

In this work, we have explored the thermodynamic properties of asymptotically flat $(\Lambda = 0)$ black holes in four-dimensional Einstein-Gauss-Bonnet (4D-EGB) gravity by employing Rényi entropy instead of the conventional Bekenstein-Hawking entropy formulation. The substitution introduces a non-extensive entropy framework that is well-suited for gravitational systems with long-range interactions or deviations from standard thermodynamic behaviour. The resulting formulation along with a restricted-phase-space-thermodynamics (RPST) like formalism, has been shown to exhibit a remarkable structural parallel with that of the RPST for the case of AdS $(\Lambda \neq 0)$ black hole.\\

A central theme of our study is the emergence of a novel thermodynamic conjugate pair: the inverse Rényi parameter $\beta$, which encodes the deviation from extensivity, and its corresponding response potential $\zeta$. This pair plays a role analogous to the central charge and chemical potential in RPST, and their thermodynamic behavior under different ensembles (fixed $\tilde{Q}$ and fixed $\tilde{\Phi}$ ensembles) closely mirrors that of which is found in holographic AdS black hole systems. Notably, we observe Van der Waals-like first-order phase transitions in both ensembles, which is a surprising result for asymptotically flat black holes that do not possess a cosmological constant. These transitions typically only appear in AdS setups, and their emergence here is attributed entirely to the non-extensive nature of the Rényi entropy.\\

We also examined the $\zeta-\beta$ processes in both fixed-$\tilde{Q}$ and fixed-$\tilde{\Phi}$ ensembles, and found that their behaviour strikingly resembles that of the $\mu-C$ processes in AdS RPST thermodynamics. This observation adds further weight to the interpretation that the inverse Rényi parameter, $\beta$ plays a role similar to the central charge, $C$ of the dual conformal field theory in holographic constructions. Moreover, a previously proposed relation between the Rényi parameter and the cosmological constant hints at a deeper link between flat-space thermodynamics and AdS holography, a connection which our results help substantiate.\\

An important component of our study is the analysis of the system's thermodynamic geometry using the framework of geometrothermodynamics (GTD). We constructed the GTD metrics for both fixed-$\tilde{Q}$ and fixed-$\tilde{\Phi}$ ensembles and compared the resulting geometries with those obtained in AdS black hole thermodynamics under the RPST formalism. Our results demonstrate that the GTD geometries of the Rényi-modified flat 4D-EGB black hole and the AdS black hole in RPST are both qualitatively similar. In addition to that we also did the topological analysis of the black hole system for both  fixed-$\tilde{Q}$ and fixed-$\tilde{\Phi}$ ensembles, and the results there revealed that both the black hole systems share the same topological class (i.e. $W=+1$) across all ensembles. Thus the geometrical along with the topological concordance reinforces the thermodynamic equivalence between the two systems, even in the absence of a cosmological constant.\\

Taken together, these findings suggest that the combination of Rényi entropy and modified gravity theories such as the Einstein-Gauss-Bonnet gravity can effectively reproduce many features of AdS thermodynamics, including critical phenomena and thermodynamic dualities. This not only expands the applicability of the RPST paradigm beyond AdS spacetimes but also strengthens the case for using non-extensive statistical mechanics as a fundamental tool in gravitational thermodynamics.\\

Looking ahead, the potential connections between the Rényi parameter and holographic quantities such as the central charge invite further exploration, particularly in the context of emergent space-time and quantum information geometry. Our study opens up the possibility that holographic thermodynamic structures are not exclusive to AdS backgrounds, but may arise more generally in gravitational systems when entropy is appropriately generalized.

\section{Acknowledgments}
	The authors would like to thank Bidyut Hazarika for numerous enlightening discussions and certain exquisite suggestions he offered during the course of this work.


\begin{thebibliography}{99}

\bibitem{Bekenstein1973}
J. D. Bekenstein, ``Black holes and entropy,'' \emph{Phys. Rev. D} \textbf{7}, 2333 (1973).

\bibitem{Hawking1975}
S. W. Hawking, ``Particle creation by black holes,'' \emph{Commun. Math. Phys.} \textbf{43}, 199 (1975).

\bibitem{Bardeen1973}
J. M. Bardeen, B. Carter, and S. W. Hawking, ``The four laws of black hole mechanics,'' \emph{Commun. Math. Phys.} \textbf{31}, 161 (1973).

\bibitem{Bamba2012}
K. Bamba, R. Myrzakulov, S. Nojiri, and S. D. Odintsov, ``Reconstruction of $f(T)$ gravity: Rip cosmology, finite-time future singularities and thermodynamics,'' \emph{Phys. Rev. D} \textbf{85}, 104036 (2012), [arXiv:1202.4057 [gr-qc]].


\bibitem{Nojiri2011}
S. Nojiri and S. D. Odintsov, ``Unified cosmic history in modified gravity: from F(R) theory to Lorentz non-invariant models,'' \emph{Phys. Rept.} \textbf{505}, 59–144 (2011).

\bibitem{Page2005}
D. N. Page, ``Hawking radiation and black hole thermodynamics,'' \emph{New J. Phys.} \textbf{7}, 203 (2005).

\bibitem{Carlip2014}
S. Carlip, ``Black hole thermodynamics,'' \emph{Int. J. Mod. Phys. D} \textbf{23}, 1430023 (2014).

\bibitem{Witten1998}
E. Witten, ``Anti-de Sitter space and holography,'' \emph{Adv. Theor. Math. Phys.} \textbf{2}, 253–291 (1998).

\bibitem{Hubeny2010}
V. E. Hubeny and M. Rangamani, ``A holographic view on physics out of equilibrium,'' \emph{Adv. High Energy Phys.} \textbf{2010}, 297916 (2010).

\bibitem{Rovelli2004}
C. Rovelli, \emph{Quantum Gravity}, Cambridge University Press (2004).

\bibitem{Ashtekar2005}
A. Ashtekar and M. Bojowald, ``Black hole evaporation: A paradigm,'' \emph{Class. Quant. Grav.} \textbf{22}, 3349–3362 (2005).

\bibitem{Kastor2009}
D. Kastor, S. Ray, and J. Traschen, 
``Enthalpy and the Mechanics of AdS Black Holes,'' 
\emph{Class. Quant. Grav.} \textbf{26}, 195011 (2009), 
[arXiv:0904.2765 [hep-th]].

\bibitem{Kubiznak2012}
D. Kubizňák and R. B. Mann, 
``P–V criticality of charged AdS black holes,'' 
\emph{JHEP} \textbf{1207}, 033 (2012), 
[arXiv:1205.0559 [hep-th]].

\bibitem{Kubiznak2017}
D. Kubizňák, R. B. Mann, and M. Teo, 
``Black hole chemistry: thermodynamics with Lambda,'' 
\emph{Class. Quant. Grav.} \textbf{34}, 063001 (2017), 
[arXiv:1608.06147 [hep-th]].

\bibitem{Maldacena1998}
J. M. Maldacena, 
``The Large N limit of superconformal field theories and supergravity,'' 
\emph{Adv. Theor. Math. Phys.} \textbf{2}, 231 (1998), 
[arXiv:hep-th/9711200].

\bibitem{Cong2021}
W. Cong, D. Kubizňák, and R. B. Mann, 
``Thermodynamics of AdS Black Holes: Central Charge Criticality,'' 
\emph{Phys. Rev. Lett.} \textbf{127}, 091301 (2021), 
[arXiv:2105.02223 [hep-th]].

\bibitem{Visser2022}
M. R. Visser, W. Cong, D. Kubizňák, and R. B. Mann, 
``Holographic thermodynamics requires a chemical potential for color,'' 
\emph{SciPost Phys.} \textbf{13}, 007 (2022), 
[arXiv:2203.08832 [hep-th]].

\bibitem{wang2022black}
T.~Wang and L.~Zhao,
``Black hole thermodynamics is extensive with variable Newton constant,''
\textit{Phys. Lett. B}, vol.~827, p.~136935, 2022.

\bibitem{huang2024thermodynamics}
B.-H.~Huang, H.-W.~Hu, and L.~Zhao,
``Thermodynamics for regular black holes as intermediate thermodynamic states and quasinormal frequencies,''
\textit{J. Cosmol. Astropart. Phys.}, vol.~2024, no.~03, p.~053, 2024.

\bibitem{kong2023restricted}
X.~Kong, Z.~Zhang, and L.~Zhao,
``Restricted phase space thermodynamics of charged AdS black holes in conformal gravity,''
\textit{Chin. Phys. C}, vol.~47, no.~9, p.~095105, 2023.

\bibitem{hazarika2025rpst}
B.~Hazarika and P.~Phukon,  
``RPST-inspired formalism for black holes in flat spacetime,''  
\textit{Phys. Lett. B}, p.~139477, 2025.

\bibitem{Lovelock1971}
D. Lovelock, 
``The Einstein tensor and its generalizations,'' 
\emph{J. Math. Phys.} \textbf{12}, 498 (1971).

\bibitem{Lanczos1938}
C. Lanczos, 
``A remarkable property of the Riemann-Christoffel tensor in four dimensions,'' 
\emph{Annals Math.} \textbf{39}, 842 (1938).










\bibitem{kumar2020hayward}
A.~Kumar, D.~V.~Singh, and S.~G.~Ghosh,
``Hayward black holes in Einstein--Gauss--Bonnet gravity,''
\textit{Annals of Physics}, vol.~419, p.~168214, 2020.

\bibitem{shahraeini2022radiation}
S.~S.~Shahraeini, K.~Nozari, and S.~Saghafi,
``Radiation from Hayward black hole via tunneling process in Einstein–Gauss–Bonnet gravity,''
\textit{J. Hologr. Appl. Phys.}, vol.~2, no.~4, pp.~55--62, 2022.

\bibitem{glavan2020einstein}
D.~Glavan and C.~Lin,
``Einstein–Gauss–Bonnet gravity in four-dimensional spacetime,''
\textit{Phys. Rev. Lett.}, vol.~124, no.~8, p.~081301, 2020.

\bibitem{motohashi2015third}
Motohashi, Hayato and Suyama, Teruaki,
\textit{Third order equations of motion and the Ostrogradsky instability},
Phys. Rev. D \textbf{91}, 085009 (2015).

\bibitem{fernandes2020charged}
P.~G.~S.~Fernandes,
``Charged black holes in AdS spaces in 4D Einstein Gauss–Bonnet gravity,''
\textit{Phys. Lett. B}, vol.~805, p.~135468, 2020.

\bibitem{wei2020extended}
S.~W.~Wei and Y.~X.~Liu,
``Extended thermodynamics and microstructures of four-dimensional charged Gauss–Bonnet black hole in AdS space,''
\textit{Phys. Rev. D}, vol.~101, no.~10, p.~104018, 2020.

\bibitem{mansoori2021thermodynamic}
S.~A.~H.~Mansoori,
``Thermodynamic geometry of the novel 4-D Gauss–Bonnet AdS black hole,''
\textit{Phys. Dark Universe}, vol.~31, p.~100776, 2021.

\bibitem{bousder2021particle}
M.~Bousder and M.~Bennai,
``Particle–antiparticle in 4D charged Einstein–Gauss–Bonnet black hole,''
\textit{Phys. Lett. B}, vol.~817, p.~136343, 2021.

\bibitem{guo2020innermost}
M.~Guo and P.~C.~Li,
``Innermost stable circular orbit and shadow of the 4D Einstein–Gauss–Bonnet black hole,''
\textit{Eur. Phys. J. C}, vol.~80, no.~6, pp.~1--8, 2020.

\bibitem{vagnozzi2023horizon}
S.~Vagnozzi, R.~Roy, Y.~D.~Tsai, L.~Visinelli, M.~Afrin, A.~Allahyari, P.~Bambhaniya, D.~Dey, S.~G.~Ghosh, P.~S.~Joshi, \textit{et al.},
``Horizon-scale tests of gravity theories and fundamental physics from the Event Horizon Telescope image of Sagittarius A,''
\textit{Class. Quant. Grav.}, vol.~40, no.~16, p.~165007, 2023.

\bibitem{zhang2020greybody}
C.~Y.~Zhang, P.~C.~Li, and M.~Guo,
``Greybody factor and power spectra of the Hawking radiation in the 4D Einstein–Gauss–Bonnet de Sitter gravity,''
\textit{Eur. Phys. J. C}, vol.~80, no.~9, p.~874, 2020.

\bibitem{zhang2020superradiance}
C.~Y.~Zhang, S.~J.~Zhang, P.~C.~Li, and M.~Guo,
``Superradiance and stability of the regularized 4D charged Einstein–Gauss–Bonnet black hole,''
\textit{J. High Energy Phys.}, vol.~2020, no.~8, pp.~1--19, 2020.

\bibitem{bravo2022nonlinear}
M.~Bravo-Gaete, L.~Guajardo, and J.~Oliva,
``Nonlinear charged planar black holes in four-dimensional scalar–Gauss–Bonnet theories,''
\textit{Phys. Rev. D}, vol.~106, no.~2, p.~024017, 2022.

\bibitem{yang2020weak}
S.~J.~Yang, J.~J.~Wan, J.~Chen, J.~Yang, and Y.~Q.~Wang,
``Weak cosmic censorship conjecture for the novel 4D charged Einstein–Gauss–Bonnet black hole with test scalar field and particle,''
\textit{Eur. Phys. J. C}, vol.~80, pp.~1--11, 2020.























\bibitem{Renyi1970}
A. Rényi, 
\textit{Probability Theory} 
(North-Holland, Amsterdam, 1970).

\bibitem{Tsallis2013}
C. Tsallis, 
``Nonadditive entropy and nonextensive statistical mechanics: An overview after 20 years,'' 
\emph{Braz. J. Phys.} \textbf{39}, 337–356 (2009), 
[arXiv:0812.4370 [cond-mat.stat-mech]].

\bibitem{Biro2013}
T. S. Biró and V. G. Czinner, 
``A q-parameter bound for particle spectra based on black hole thermodynamics with Rényi entropy,'' 
\emph{Phys. Lett. B} \textbf{726}, 861–865 (2013), 
[arXiv:1309.4261 [gr-qc]].

\bibitem{Czinner2016}
V. G. Czinner and H. Iguchi, 
``Rényi entropy and the thermodynamic stability of black holes,'' 
\emph{Phys. Lett. B} \textbf{752}, 306–310 (2016), 
[arXiv:1511.06963 [gr-qc]].

\bibitem{Barzi1}
F. Barzi and H. El Moumni, 
``On Rényi universality formula of charged flat black holes from Hawking-Page phase transition,'' 
\emph{Phys. Lett. B} \textbf{833}, 137378 (2022), 
doi:10.1016/j.physletb.2022.137378, 
[arXiv:2209.08195 [hep-th]].

\bibitem{Barzi2}
F. Barzi, H. El Moumni and K. Masmar, 
``Riemann Surfaces and Winding Numbers of Rényi Phase Structure of Charged-Flat Black Holes,'' 
\emph{Eur. Phys. J. C} \textbf{84}, 1141 (2024), 
doi:10.1140/epjc/s10052-024-13511-0, 
[arXiv:2408.05870 [hep-th]].

\bibitem{proof3}
E. Hirunsirisawat, R. Nakarachinda and C. Promsiri, 
``Emergent phase, thermodynamic geometry, and criticality of charged black holes from Rényi statistics,'' 
\emph{Phys. Rev. D} \textbf{105}, 124049 (2022), 
doi:10.1103/PhysRevD.105.124049, 
[arXiv:2204.13023 [hep-th]].

\bibitem{proof4}
F. Barzi, H. El Moumni and K. Masmar, 
``On some phase equilibrium features of charged black holes in flat spacetime via Rényi statistics,'' 
\emph{Gen. Rel. Grav.} \textbf{55}, 109 (2023), 
doi:10.1007/s10714-023-03158-9, 
[arXiv:2304.04945 [gr-qc]].

\bibitem{proof5}
P. Chunaksorn, E. Hirunsirisawat, R. Nakarachinda, L. Tannukij and P. Wongjun, 
``Thermodynamics of asymptotically de Sitter black hole in dRGT massive gravity from Rényi entropy,'' 
\emph{Eur. Phys. J. C} \textbf{82}, 1174 (2022), 
doi:10.1140/epjc/s10052-022-11110-5, 
[arXiv:2208.14770 [gr-qc]].

\bibitem{proof6}
Z. Wang, H. Ren, J. Chen and Y. Wang, 
``Thermodynamics and phase transition of Bardeen black hole via Rényi statistics in grand canonical ensemble and canonical ensemble,'' 
\emph{Eur. Phys. J. C} \textbf{83}, 527 (2023), 
doi:10.1140/epjc/s10052-023-11680-y.

\bibitem{proof7}
F. Barzi, H. El Moumni and K. Masmar, 
``Rényi topology of charged-flat black hole: Hawking-Page and Van-der-Waals phase transitions,'' 
\emph{JHEAp} \textbf{42}, 63–86 (2024), 
doi:10.1016/j.jheap.2024.03.005, 
[arXiv:2309.14069 [hep-th]].

\bibitem{proof8}
F. Barzi, H. El Moumni and K. Masmar, 
``Thermal chaos of charged-flat black hole via Rényi formalism,'' 
\emph{Nucl. Phys. B} \textbf{1005}, 116606 (2024), 
doi:10.1016/j.nuclphysb.2024.116606, 
[arXiv:2404.14609 [hep-th]].

\bibitem{proof9}
A. Baruah and P. Phukon, 
``Restricted Phase Space Thermodynamics of Dyonic AdS Black Holes: Comparative Analysis Using Different Entropy Models,'' 
[arXiv:2411.02273 [hep-th]].

\bibitem{expanding}
H. R. Fazlollahi, 
``Rényi entropy correction to expanding universe,'' 
\emph{Eur. Phys. J. C} \textbf{83}, 29 (2023), 
doi:10.1140/epjc/s10052-023-11183-w.

\bibitem{Quevedo2007}
H. Quevedo, 
``Geometrothermodynamics,'' 
\emph{J. Math. Phys.} \textbf{48}, 013506 (2007), 
doi:10.1063/1.2409524.

\bibitem{Quevedo2008}
H. Quevedo, 
``Geometrothermodynamics of black holes,'' 
\emph{Gen. Rel. Grav.} \textbf{40}, 971–984 (2008), 
doi:10.1007/s10714-007-0586-0.

\bibitem{Soroushfar2016}
S. Soroushfar, R. Saffari and N. Kamvar, 
``Thermodynamic geometry of black holes in \( f(R) \) gravity,'' 
\emph{Eur. Phys. J. C} \textbf{76}, 476 (2016), 
doi:10.1140/epjc/s10052-016-4311-8.






\bibitem{Wei2024}
S.-W. Wei, Y.-X. Liu, and R. B. Mann,
``Universal topological classifications of black hole thermodynamics,''
\emph{Phys. Rev. D} \textbf{110}, L081501 (2024),
[arXiv:2409.09333 [gr-qc]].

\bibitem{Wu2025}
D. Wu, W. Liu, S.-Q. Wu, and R. B. Mann,
``Novel topological classes in black hole thermodynamics,''
\emph{Phys. Rev. D} \textbf{111}, L061501 (2025),
[arXiv:2411.10102 [hep-th]].

\bibitem{Wu2024}
D. Wu, S. Gu, X. Zhu, \emph{et al.},
``Universal thermodynamic topological classes of rotating black holes,''
\emph{JHEP} \textbf{06}, 213 (2024),
[arXiv:2409.12747 [gr-qc]].

\bibitem{Wei2022PRL}
S.-W. Wei, Y.-X. Liu, and R. B. Mann,
``Black hole solutions as topological thermodynamic defects,''
\emph{Phys. Rev. Lett.} \textbf{129}, 191101 (2022),
[arXiv:2208.01932 [gr-qc]].

\bibitem{Wu2023a}
D. Wu,
``Topological classes of rotating black holes,''
\emph{Phys. Rev. D} \textbf{107}, 024024 (2023),
[arXiv:2211.15151 [gr-qc]].

\bibitem{Wu2023b}
D. Wu and S.-Q. Wu,
``Topological classes of thermodynamics of rotating AdS black holes,''
\emph{Phys. Rev. D} \textbf{107}, 084002 (2023),
[arXiv:2301.03002 [gr-qc]].

\bibitem{Wei2022PRD}
S.-W. Wei and Y.-X. Liu,
``Topology of black hole thermodynamics,''
\emph{Phys. Rev. D} \textbf{105}, 104003 (2022),
[arXiv:2112.01706 [gr-qc]].

\bibitem{Schouten1951}
J. A. Schouten, 
\textit{Tensor Analysis for Physics} 
(Oxford at the Clarendon Press, 1951).













\bibitem{FernandesReview2022}
P.~G.~S.~Fernandes, P.~Carrilho, T.~Clifton and D.~J.~Mulryne,
``The 4D Einstein-Gauss-Bonnet theory of gravity: a review,''
Class. Quant. Grav. \textbf{39}, no.6, 063001 (2022),
doi:10.1088/1361-6382/ac500a
[arXiv:2201.03060 [gr-qc]].

\bibitem{Ladghami2023}
Y.~Ladghami, B.~Asfour, A.~Bouali, A.~Errahmani and T.~Ouali,
``4D-EGB black holes in RPS thermodynamics,''
Phys. Dark Univ. \textbf{41}, 101261 (2023),
doi:10.1016/j.dark.2023.101261.






\end{thebibliography}
	\end{document}